\documentclass[pre,twocolumn,aps,showpacs]{revtex4}
\newcommand{\bec}[1]{\mbox{\boldmath $ #1$}}
\usepackage{graphicx}
\begin{document}
\bigskip
\bigskip
\title{Effects of differential and uniform rotation on nonlinear electromotive
force in a turbulent flow}
\author{Igor Rogachevskii}
\email{gary@menix.bgu.ac.il} \homepage{http://www.bgu.ac.il/~gary}
\author{Nathan Kleeorin}
\email{nat@menix.bgu.ac.il} \affiliation{Department of Mechanical
Engineering, The Ben-Gurion
University of the Negev, \\
POB 653, Beer-Sheva 84105, Israel}
\date{\today}
\begin{abstract}
An effect of the differential rotation on the nonlinear
electromotive force in MHD turbulence is found. It includes a
nonhelical $\alpha$ effect which is caused by a differential
rotation, and it is independent of a hydrodynamic helicity. There
is no quenching of this effect contrary to the quenching of the
usual $\alpha$ effect caused by a hydrodynamic helicity. The
nonhelical $\alpha$ effect vanishes when the rotation is constant
on the cylinders which are parallel to the rotation axis. The mean
differential rotation creates also the shear-current effect which
changes its sign with the nonlinear growth of the mean magnetic
field. However, there is no quenching of this effect. These
phenomena determine the nonlinear evolution of the mean magnetic
field. An effect of a uniform rotation on the nonlinear
electromotive force is also studied. A nonlinear theory of the
${\bf \Omega} {\bf \times} \bar{\bf J}$ effect is developed, and
the quenching of the hydrodynamic part of the usual $\alpha$
effect which is caused by a uniform rotation and inhomogeneity of
turbulence, is found. Other contributions of a uniform rotation to
the nonlinear electromotive force are also determined. All these
effects are studied using the spectral $\tau$ approximation (the
third-order closure procedure). An axisymmetric mean-field dynamo
in the spherical and cylindrical geometries is considered. The
nonlinear saturation mechanism based on the magnetic helicity
evolution is discussed. It is shown that this universal mechanism
is nearly independent of the form of the flux of magnetic
helicity, and it requires only a nonzero flux of magnetic
helicity. Astrophysical applications of these effects are
discussed.
\end{abstract}


\maketitle

\section{INTRODUCTION}

Generation of magnetic fields by a turbulent flow of conducting
fluid is a fundamental problem which has a large number of
applications in solar physics, astrophysics, geophysics, planetary
physics and in laboratory studies (see, e.g.,
\cite{M78,P79,KR80,ZRS83,RSS88,S89,RS92,B2000,RH04}, and
references therein). In recent time the problem of nonlinear
mean-field magnetic dynamo is a subject of active discussions
(see, e.g.,
\cite{BB96,K99,GD94,GD96,KRR95,S96,CH96,KR99,FB99,RK2000,RK2001,KMRS2000,KMRS02,BB02,BS04},
and references therein). The conventional approach to the
nonlinear dynamo is based on comparison of the three effects
participating in dynamo action, namely the $\alpha$ effect (caused
by helical motions of a turbulent fluid), the large-scale
differential (nonuniform) rotation $\delta \Omega$ and the
turbulent magnetic diffusivity $\eta_{_{T}}$. The mean magnetic
field is generated due to a combined effect of the differential
rotation and the $\alpha$ effect. These effects have been
considered as independent phenomena. In particular, the
electromotive force has been determined independently of the
differential rotation.

On the other hand, the differential rotation can be regarded as
large-scale motions with a mean velocity shear imposed on the
small-scale turbulent fluid flow. An interaction of the mean
differential rotation with the small-scale turbulent motions can
cause a generation of a mean magnetic field even in a nonhelical,
homogeneous and incompressible turbulent fluid flow. This
mechanism of mean-field dynamo is associated with a shear-current
effect which is determined by the $\bar{\bf W} {\bf \times}
\bar{\bf J}$ term in the electromotive force, where $\bar{\bf W}$
is the mean vorticity caused by the mean velocity shear and
$\bar{\bf J}$ is the mean electric current (see \cite{RK03}). A
nonlinear theory of a shear-current effect in a nonrotating
homogeneous and nonhelical turbulence with an imposed mean
velocity shear in a plane geometry was developed in \cite{RK04}.
It was shown that during the nonlinear growth of the mean magnetic
field, the shear-current effect changes its sign, but there is no
quenching of this effect contrary to the quenching of the usual
$\alpha$ effect, the nonlinear turbulent magnetic diffusion, etc.

In this study we investigated the effects of differential and
uniform rotation on nonlinear electromotive force. The main
conclusion of this study is that the nonlinear electromotive force
cannot be determined independently of the mean differential
rotation. We found a nonhelical $\alpha$ effect which is caused by
a differential rotation and is independent of a hydrodynamic
helicity. There is no quenching of this effect contrary to the
quenching of the usual $\alpha$ effect caused by a hydrodynamic
helicity. The mean differential rotation of fluid can decrease the
total $\alpha$ effect due to the nonhelical $\alpha$ effect. Two
kinds of the $\alpha$ effect (helical and nonhelical) have
opposite signs. Therefore, the total $\alpha$ effect should always
change its sign during the nonlinear growth of the mean magnetic
field because there is a quenching of the usual (helical) $\alpha$
effect. This can saturate the growth of the mean magnetic field.

The mean differential rotation creates also the shear-current
effect. We found that the mean differential rotation increases the
growth rate of the large-scale dynamo instability at a weak mean
magnetic field due to the shear-current effect, and causes a
saturation of the growth of the mean magnetic field at a stronger
field. Note that the applications of the obtained results to the
solar convective zone shows that the nonlinear shear-current
effect becomes dominant at least at the base of the convective
zone. We found that the nonlinear function $\sigma_0(\bar{B})$
defining the shear-current effect is the same for a turbulence
with a mean differential rotation in cylindrical and spherical
geometries for an axisymmetric mean field dynamo problem and for a
nonrotating turbulence with an imposed linear mean velocity shear
in a plane geometry. The latter case was investigated in
\cite{RK04}.

We also studied an effect of a uniform rotation on the nonlinear
electromotive force. In particular, we developed a nonlinear
theory of the ${\bf \Omega} {\bf \times} \bar{\bf J}$ effect and
we determined the nonlinear quenching of the hydrodynamic part of
the $\alpha$ effect which is caused by both, a uniform rotation
and inhomogeneity of turbulence. Other nonlinear coefficients
defining the nonlinear electromotive force are also determined as
a function of a uniform rotation. In this study we considered a
uniform rotation with a small rotation rate in comparison with the
correlation time of the fluid turbulent velocity field. We studied
all the above effects using the spectral $\tau$ approximation (the
third-order closure procedure).

This paper is organized as follows. In Section II we formulated
the assumptions and the method of the derivation of the nonlinear
electromotive force in a turbulence with a uniform and nonuniform
rotations. In Section III we considered axisymmetric mean-field
dynamo equations and determined the coefficients defining the
electromotive force for a rotating turbulence. In Section III we
also discussed in details the effects of differential and uniform
rotation on nonlinear coefficient defining the electromotive
force. In Section IV we analyzed the nonlinear saturation of the
mean magnetic field and discussed the astrophysical applications
of the obtained results. In Appendix A we derived the nonlinear
electromotive force in a turbulence with uniform and nonuniform
rotations.

\section{THE METHOD OF DERIVATIONS}

In a framework of the mean-field approach the evolution of the
mean magnetic field $\bar{\bf B}$ is determined by equation
\begin{eqnarray}
{\partial \bar{\bf B} \over \partial t} = \bec{\nabla} \times
(\bar{\bf U} \times \bar{\bf B} + \bec{\cal E} - \eta \bec{\nabla}
\times \bar{\bf B}) \;
\label{A12}
\end{eqnarray}
(see, e.g., \cite{M78,P79,KR80,ZRS83,RSS88,S89}), where $ \bar{\bf
U} $ is a mean velocity (the differential rotation), $ \eta $ is
the magnetic diffusion due to the electrical conductivity of
fluid. The general form of the electromotive force $ \bec{\cal E}
= \langle {\bf u} \times {\bf b} \rangle $ in an anisotropic
turbulence is given by
\begin{eqnarray}
{\cal E}_{i} &=& \alpha_{ij} \bar B_{j} + ({\bf V}^{\rm eff} {\bf
\times} \bar{\bf B})_{i} - \eta_{ij} (\bec{\nabla} {\bf \times}
\bar{\bf B})_{j}
\nonumber\\
&& - [\bec{\delta} {\bf \times} (\bec{\nabla} {\bf \times}
\bar{\bf B})]_{i} - \kappa_{ijk} (\partial \hat B)_{jk}  \;
\label{A14}
\end{eqnarray}
(see \cite{R80,RKR03}), where $(\partial \hat B)_{ij} = (1/2)
(\nabla_{i} \bar B_{j} + \nabla_{j} \bar B_{i}) ,$ ${\bf u}$ and
${\bf b}$ are fluctuations of the velocity and magnetic field,
respectively, angular brackets denote averaging over an ensemble
of turbulent fluctuations, the tensors $\alpha_{ij}$ and
$\eta_{ij}$ describe the $\alpha$-effect and the turbulent
magnetic diffusion, respectively, ${\bf V}^{\rm eff}$ is the
effective diamagnetic (or paramagnetic) velocity, $\kappa_{ijk}$
and $\bec{\delta}$ describe an evolution of the mean magnetic
field in an anisotropic turbulence. Nonlinearities in the
mean-field dynamo imply dependencies of the coefficients $
(\alpha_{ij} , \eta_{ij}, {\bf V}^{\rm eff}, $ {\em etc.})
defining the electromotive force on the mean magnetic field.

The method of the derivation of equation for the nonlinear
electromotive force in a rotating turbulence is similar to that
used in \cite{RK04} for a nonrotating turbulence with an imposed
mean velocity shear. We consider the case of large hydrodynamic
and magnetic Reynolds numbers. The momentum equation and the
induction equation for the turbulent fields in a frame rotating
with an angular velocity ${\bf \Omega}$ are given by
\begin{eqnarray}
{\partial {\bf u}(t,{\bf x}) \over \partial t} &=& - {\bec{\nabla}
p_{\rm tot} \over \rho_0} + {1 \over \mu \rho_0} [({\bf b} \cdot
\bec{\nabla}) \bar{\bf B} + (\bar{\bf B} \cdot \bec{\nabla}){\bf
b}]
\nonumber \\
&& + 2 {\bf u} {\bf \times} {\bf \Omega} - (\bar{\bf U} \cdot
\bec{\nabla}) {\bf u} - ({\bf u} \cdot \bec{\nabla}) \bar{\bf U}
\nonumber \\
&& + {\bf u}^N + {\bf F} \,,
\label{B1} \\
{\partial {\bf b}(t,{\bf x}) \over \partial t} &=& (\bar{\bf B}
\cdot \bec{\nabla}){\bf u} - ({\bf u} \cdot \bec{\nabla}) \bar{\bf
B} - (\bar{\bf U} \cdot \bec{\nabla}) {\bf b}
\nonumber \\
&& + ({\bf b} \cdot \bec{\nabla}) \bar{\bf U} + {\bf b}^N \,,
\label{B2}
\end{eqnarray}
where $\bec{\nabla} \cdot {\bf u} = 0$, $\, \rho_0$ is the fluid
density, $\mu$ is the magnetic permeability of the fluid, $\rho_0
{\bf F}$ is a random external stirring force, ${\bf u}^{N}$ and
${\bf b}^{N}$ are the nonlinear terms which include the molecular
dissipative terms, $p_{\rm tot} = p + \mu^{-1} \,(\bar{\bf B}
\cdot {\bf b}) $ are fluctuations of the total pressure, $p$ are
fluctuations of the fluid pressure. Hereafter we omit the magnetic
permeability of the fluid, $\mu$, in equations, i.e., we include
$\mu^{-1/2}$ in the definition of magnetic field. We study the
effect of a mean rotation of the fluid on the nonlinear
electromotive force. We split rotation into uniform and
differential parts. By means of Eqs.~(\ref{B1})-(\ref{B2}) written
in a Fourier space we derive equations for the correlation
functions of the velocity field $f_{ij}({\bf k}) = \hat L(u_i;
u_j)$, of the magnetic field $h_{ij}({\bf k}) = \hat L(b_i; b_j) $
and for the cross helicity $g_{ij}({\bf k}) = \hat L(b_i; u_j)$,
where
\begin{eqnarray}
\hat L(a; c) = \int \langle a(t,{\bf k} + {\bf  K} / 2) c(t,-{\bf
k} + {\bf  K} / 2) \rangle
\nonumber \\
\times \exp{(i {\bf K} {\bf \cdot} {\bf R}) } \,d {\bf  K} \;,
\label{X20}
\end{eqnarray}
and ${\bf R}$ and ${\bf K}$ correspond to the large scales, and
${\bf r}$ and ${\bf k}$ to the  small ones (see, e.g.,
\cite{RS75,KR94}). The equations for these correlation functions
are given by Eqs. (\ref{B6})-(\ref{B8}) in Appendix A. These
equations for the second moments contain high moments and a
closure problem arises (see, e.g., \cite{O70,MY75,Mc90}). We apply
the spectral $\tau$ approximation or the third-order closure
procedure (see, e.g.,
\cite{O70,PFL76,KRR90,KMR96,RK2000,RK2001,KR03,BK04}), which
allows to express the deviations of the third moments from the
background turbulence in ${\bf k}$ space in terms of the
corresponding deviations of the second moments, e.g.,
\begin{eqnarray}
\hat D f_{ij}^N - \hat D f_{ij}^{N(0)} &=& - (f_{ij} -
f_{ij}^{(0)}) / \tau (k) \;,
\label{A1} \\
\hat D h_{ij}^{N} - \hat D h_{ij}^{N(0)} &=& - (h_{ij} -
h_{ij}^{(0)}) / \tau(k) \;,
\label{A2} \\
\hat D g_{ij}^N &=& - g_{ij} / \tau (k) \;, \label{A3}
\end{eqnarray}
where the tensors $ \hat D f_{ij}^N ,$ $ \, \hat D h_{ij}^N $ and
$ \hat D g_{ij}^N $ are related to the third moments in equations
for the second moments $ f_{ij}, h_{ij} $ and $ g_{ij} ,$
respectively (see Eqs. (\ref{B6})-(\ref{B8}) in Appendix A). The
correlation functions with the superscript $ {(0)} $ determine the
background turbulence (with a zero mean magnetic field, $ \bar
{\bf B} = 0),$ and $ h_{ij}^{(0)} $ is the nonhelical part of the
tensor of magnetic fluctuations of the background turbulence, $
\tau (k) $ is the characteristic relaxation time of the
statistical moments. We applied the $ \tau $-approximation only
for the nonhelical part $h_{ij}$ of the tensor of magnetic
fluctuations. The helical part $h_{ij}^{(H)}$ depends on the
magnetic helicity, and it is determined by the dynamic equation
which follows from the magnetic helicity conservation arguments
\cite{KR82,ZRS83} (see also
\cite{GD94,KRR95,KR99,KMRS2000,KMRS02,BB02}). In the present paper
we consider an intermediate nonlinearity which implies that the
mean magnetic field is not enough strong in order to affect the
correlation time of turbulent velocity field. We also consider
uniform rotation with a small rotation rate in comparison with the
correlation time of the fluid turbulent velocity field $(\Omega \,
\tau_0 \ll 1). $ The mean velocity shear due to the differential
rotation is considered to be weak $(\delta \Omega \, \tau_0 \ll
1). $ For the integration in $ {\bf k} $-space of the second
moments we use the following model of the background turbulence
(with zero mean magnetic field, $ \bar{\bf B} = 0$ and without
rotation):
\begin{eqnarray}
f_{ij}^{(0)}({\bf k}) &=& E(k) \biggl\{\langle {\bf u}^2
\rangle^{(0)} \biggl[\delta_{ij} - k_{ij} + {i \over 2 k^2} (k_i
\Lambda_j^{(v)}
\nonumber \\
& & - k_j \Lambda_i^{(v)}) \biggr] - \frac{1}{2 k^2}
\biggl[\varepsilon_{ijn} (2 i k_n + k_{mn} \nabla_m)
\nonumber \\
& & - (k_{in} \varepsilon_{jnm} + k_{jn} \varepsilon_{inm})
\nabla_m \biggr] \mu^{v} \biggr\}\;,
\label{K1}\\
h_{ij}^{(0)}({\bf k}) &=& \langle {\bf b}^2 \rangle^{(0)} E(k)
\biggl[\delta_{ij} - k_{ij} + {i \over 2 k^2} (k_i \Lambda_j^{(b)}
\nonumber \\
& & - k_j \Lambda_i^{(b)})\biggr] \;,
\label{K2}
\end{eqnarray}
where $\varepsilon_{ijk}$ is the Levi-Civita tensor, $\delta_{ij}$
is the Kronecker tensor, $k_{ij} = k_{i} k_{j} / k^{2} ,$ $\, E(k)
= - (d \bar \tau(k) / dk) / 8 \pi k^{2} ,$ $ \, \tau(k) = 2
\tau_{0} \bar \tau(k) ,$ $ \, \bar \tau(k) = (k / k_{0})^{1-q} ,$
$\, 1 < q < 3 $ is the exponent of the kinetic energy spectrum
(e.g., $ q = 5/3 $ for Kolmogorov spectrum), $ k_{0} = 1 / l_{0}
,$ and $ l_{0} $ is the maximum scale of turbulent motions, $
\tau_{0} = l_{0} / u_{0} $, $\, u_{0} $ is the characteristic
turbulent velocity in the scale $l_{0}$, $\, \Lambda_i^{(v)} =
\nabla_i \langle {\bf u}^2 \rangle^{(0)} / \langle {\bf u}^2
\rangle^{(0)} ,$ $\, \Lambda_i^{(b)} = \nabla_i \langle {\bf b}^2
\rangle^{(0)} / \langle {\bf b}^2 \rangle^{(0)} ,$ and $\mu^{v} =
\langle {\bf u} \cdot (\bec{\nabla} {\bf \times} {\bf u})
\rangle^{(0)}$ is the hydrodynamic helicity of the background
turbulence, $\int f_{ij}^{(0)}({\bf k}) \,d {\bf k} = (\langle
{\bf u}^2 \rangle^{(0)} / 3) \delta_{ij} $ and $\int
h_{ij}^{(0)}({\bf k}) \,d {\bf k} = (\langle {\bf b}^2
\rangle^{(0)} / 3) \delta_{ij} .$ Note that $ g_{ij}^{(0)}({\bf
k}) = 0 .$ Here we neglected a very small magnetic helicity in the
background turbulence. However, the magnetic helicity in a
turbulence with a nonzero mean magnetic field is not small (see
Section III-D). The derived equations allow us to determine the
nonlinear electromotive force $ {\cal E}_{i} = \varepsilon_{imn}
\int g_{nm}({\bf k}) \,d {\bf k} $ in a rotating turbulence (see
for details, Appendix A).

\section{THE NONLINEAR ELECTROMOTIVE FORCE IN A ROTATING
TURBULENCE FOR AN AXISYMMETRIC DYNAMO}

We consider the axisymmetric $\alpha \Omega$-dynamo problem. In
cylindrical coordinates $(\rho, \varphi, z)$ the axisymmetric mean
magnetic field, $\bar{\bf B} = B(\rho,z) \, {\bf e}_\varphi +
\bec{\nabla} {\bf \times} [A(\rho,z){\bf e}_\varphi]$, is
determined by the dimensionless equations
\begin{eqnarray}
{\partial A \over \partial t} &=& \biggl[\alpha(\bar{\bf B}) +
W_\ast \, \sigma_1(\bar{\bf B}) \, \nabla_z (\delta \Omega)\biggr]
\, B + \eta_{_{A}}(\bar{\bf B}) \Delta_s A
\nonumber \\
&& - {1 \over \rho} ({\bf V}_{A}(\bar{\bf B}) \cdot \bec{\nabla})
(\rho \, A) - W_\ast \, \sigma_0(\bar{\bf B}) \, (\hat \Omega B)
\nonumber \\
&& - \Omega_\ast \, \delta_0^\Omega(\bar{\bf B}) \,
(\bec{\hat{\omega}} \cdot \bec{\nabla}) B \;,
\label{L8} \\
{\partial B \over \partial t} &=& D \, (\hat \Omega A) + \rho
\bec{\nabla} \cdot \biggl[{1 \over \rho^2} [\eta_{_{B}}
\bec{\nabla} - {\bf V}_{B}(\bar{\bf B})] (\rho \, B) \biggr] \;,
\nonumber \\
\label{L9}
\end{eqnarray}
where $\Omega_\ast = 3 \, \Omega \, \tau_0 / R_\alpha = [\Omega /
(\delta\Omega)_\ast] \, W_\ast $ , $\, W_\ast = (l_0 / L)^2 \,
(R_\omega / R_\alpha)$, $\, \bec{\hat{\omega}} = {\bf \Omega} /
\Omega$ and
\begin{eqnarray*}
(\hat \Omega B) &=& [\nabla_z (\delta \Omega) \nabla_\rho -
\nabla_\rho (\delta \Omega) \nabla_z] (\rho \, B) \;,
\\
{\bf V}_{A}(\bar{\bf B}) &=& {\bf V}_{d}(\bar{B}) -
{\phi_3(\bar{B}) \over 2} {\bf \Lambda}^{(B)} - {\phi_2(\bar{B})
\over \rho}  {\bf e}_{\rho} \;,
\\
{\bf V}_{B}(\bar{\bf B}) &=& {\bf V}_{d}(\bar{B}) +
{\phi_2(\bar{B}) + \phi_3(\bar{B}) \over \rho} {\bf e}_{\rho} \;,
\\
{\bf V}_{d}(\bar{B}) &=& - {\phi_1(\bar{B}) \over 2}  ({\bf
\Lambda}^{(v)} - \epsilon {\bf \Lambda}^{(b)}) \, {L \over
L_{_{T}}} \;,
\end{eqnarray*}
and  $\Delta_s = \Delta - 1/\rho^2 $, and  $\, {\bf \Lambda}^{(B)}
= (\bec{\nabla} \bar{\bf B}^2) / \bar{\bf B}^2 $. The nonlinear
coefficients $\alpha(\bar{\bf B})$, $ \, \eta_{_{A}}(\bar{B}) ,$
$\, \eta_{_{B}}(\bar{B}) $ defining the nonlinear $\alpha$ effect
and the nonlinear turbulent magnetic diffusion of the poloidal and
toroidal components of the mean magnetic field, are determined by
Eqs.~(\ref{C4}) and (\ref{L3}) in Section III-A. The nonlinear
coefficients $\sigma_0(\bar{B})$ defining the shear-current effect
and $\sigma_1(\bar{B})$ defining the nonhelical $\alpha$ effect,
are determined in Section III-B. The coefficient
$\delta_0^\Omega(\bar{B})$ defining the nonlinear ${\bf \Omega}
{\bf \times} \bar{\bf J}$ effect, is determined in Section III-C.
The quenching functions $\phi_n(\bar{B})$ are determined by
Eqs.~(\ref{LLL3}) in Section III-A. Note that in the equations for
the nonlinear effective drift velocities ${\bf V}_{A}(\bar{\bf
B})$ and ${\bf V}_{B}(\bar{\bf B})$ of the poloidal and toroidal
components of the mean magnetic field we neglected small
contributions $\sim O[(l_0 / L)^2]$ caused by the mean
differential rotation.

Equations~(\ref{L8}) and (\ref{L9}) are written in the
dimensionless form, where length is measured in units of $L$, time
in units of $ L^{2} / \eta_{_{T}} $ and the mean magnetic field
$\bar{B}$ is measured in units of the equipartition energy $\bar
B_{\rm eq} = \sqrt{\rho_0} \, u_0 $, the magnetic potential $A$ is
measured in units of $R_\alpha \, L \, \bar B_{\rm eq}$, the
nonlinear $\alpha$  is measured in units of $ \alpha_\ast $ (the
maximum value of the hydrodynamic part of the $ \alpha $ effect),
the basic scale of the turbulent motions $l$ and turbulent
velocity $\sqrt{\langle {\bf u}^2 \rangle}$ at the scale $l$ are
measured in units of their maximum values $l_0$ and $u_{0}$,
respectively, the dimensionless parameters $\Lambda^{(v)}$ and
$\Lambda^{(b)}$ are measured in the units of $L^{-1}_{_{T}}$ and
$\Lambda^{(B)}$ is measured in the units of $L^{-1}$, the
differential rotation $\delta\Omega$ is measured in units of
$(\delta\Omega)_\ast$, the nonlinear turbulent magnetic diffusion
coefficients $\eta_{_{A,B}}(\bar{B})$ are measured in the units of
$\eta_{_{T}}$ and the nonlinear effective drift velocities ${\bf
V}_{A,B}(\bar{B})$ are measured in the units of $ \eta_{_{T}} / L
$. We define $R_\alpha = L \alpha_\ast / \eta_{_{T}} ,$ $\,
R_\omega = (\delta\Omega)_\ast \, L^2/\eta_{_{T}}$, the
characteristic value of the turbulent magnetic diffusivity
$\eta_{_{T}} = l_0 u_{0} / 3$, the dynamo number $D = R_\omega
R_\alpha $ and $ {\rm Rm} = l_0 u_{0} / \eta $ is the magnetic
Reynolds number.

In spherical coordinates $(r, \theta, \varphi)$ the axisymmetric
mean magnetic field, $\bar{\bf B} = B(r,\theta) \, {\bf e}_\varphi
+ \bec{\nabla} {\bf \times} [A(r,\theta){\bf e}_\varphi]$, is
determined by the dimensionless equations
\begin{eqnarray}
{\partial A \over \partial t} &=& \biggl[\alpha(\bar{\bf B}) +
W_\ast \, \sigma_1(\bar{\bf B}) \, \nabla_z (\delta \Omega)\biggr]
\, B + \eta_{_{A}}(\bar{\bf B}) \Delta_s A
\nonumber \\
&& - {1 \over r \sin \theta} ({\bf V}_{A}(\bar{\bf B}) \cdot
\bec{\nabla}) \tilde A - W_\ast \, \sigma_0(\bar{\bf B}) \, (\hat
\Omega B)
\nonumber \\
&& - \Omega_\ast \, \delta_0^\Omega(\bar{\bf B}) \,
(\bec{\hat{\omega}} \cdot \bec{\nabla}) B \;,
\label{L10} \\
{\partial B \over \partial t} &=& D \, (\hat \Omega A) + r \, \sin
\theta \, \bec{\nabla} \cdot \biggl[{1 \over r^2 \, \sin^2 \theta}
[\eta_{_{B}} \bec{\nabla}
\nonumber \\
&& - {\bf V}_{B}(\bar{\bf B})] \tilde B \biggr] \;, \label{L11}
\end{eqnarray}
where $ \tilde A = r \, \sin \theta \, A $, $\, \tilde B = r \,
\sin \theta \, B $,
\begin{eqnarray*}
(\hat \Omega B) &=& [\nabla_r (\delta \Omega) \, \nabla_\theta -
\nabla_\theta (\delta \Omega) \, \nabla_r] \tilde B \;,
\\
\nabla_z  &=& \cos \theta \, \nabla_r  - \sin \theta \,
\nabla_\theta  \;,
\\
{\bf V}_{A}(\bar{\bf B}) &=& {\bf V}_{d}(\bar{B}) -
{\phi_3(\bar{B}) \over 2} {\bf \Lambda}^{(B)} - {\phi_2(\bar{B})
\over r} \, ({\bf e}_{r} + \cot \theta \, {\bf e}_{\theta}) ,
\\
{\bf V}_{B}(\bar{\bf B}) &=& {\bf V}_{d}(\bar{B}) +
{\phi_2(\bar{B}) + \phi_3(\bar{B}) \over r} \, ({\bf e}_{r} + \cot
\theta \, {\bf e}_{\theta}) \;,
\end{eqnarray*}
$\Delta_s = \Delta - 1 / (r \, \sin \theta)^2 $ and $\nabla_\theta
= (1/r) \, (\partial / \partial \theta) $. Note that $\rho = r \,
\sin \theta$.

\subsection{The nonlinear $\alpha$ effect and the nonlinear
turbulent magnetic diffusion  coefficients of the mean
magnetic field}

The nonlinear $\alpha$ effect is given by $\alpha(\bar{\bf B}) =
\alpha^v + \alpha^m ,$ where $\alpha^v = \chi^v \phi^v(\bar{B}) +
\alpha^\Omega $ is the hydrodynamic part of the $\alpha$ effect,
and $\alpha^m = \chi^c(\bar{\bf B}) \phi^m(\bar{B}) $ is the
magnetic part of the $\alpha$ effect, and the dimensionless
parameter $\chi^v = - \tau_0 \, \mu^{v} / 3 \alpha_\ast$ is
related to the hydrodynamic helicity $\mu^{v} = \langle {\bf u}
\cdot (\bec{\nabla} {\bf \times} {\bf u}) \rangle^{(0)}$ of the
background turbulence, the dimensionless function $
\chi^{c}(\bar{\bf B}) = (\tau_0 / 3 \rho_0 \alpha_\ast) \langle
{\bf b} \cdot (\bec{\nabla} {\bf \times} {\bf b}) \rangle $ is
related to the current helicity $\langle {\bf b} \cdot
(\bec{\nabla} {\bf \times} {\bf b}) \rangle$. Here $ \chi^v $ and
$ \chi^c$ are measured in units of $ \alpha_\ast $, $\, \tau_0 =
l_0 / u_{0}$ is the correlation time of turbulent velocity field
and $\alpha^\Omega$ is the contribution to the hydrodynamic part
of the $\alpha$ effect caused by a uniform rotation and
inhomogeneity of turbulence. Thus,
\begin{eqnarray}
\alpha(\bar{\bf B}) = \chi^v \phi^v(\bar{B}) + \alpha^\Omega +
\chi^c(\bar{\bf B}) \phi^m(\bar{B}) \;, \label{C4}
\end{eqnarray}
[see Eqs.~(\ref{BL1}) and~(\ref{CC4}) in Appendix A], where
\begin{eqnarray}
\alpha^\Omega &=& - {2 \over 3} \, {L \, \Omega_\ast \over
L_{_{T}}} \, \bec{\hat{\omega}} {\bf \cdot}
[\phi^\Omega_1(\bar{B}) \, {\bf \Lambda}^{(v)} + \epsilon \,
\phi^\Omega_2(\bar{B}) \, {\bf \Lambda}^{(b)}] \;,
 \label{CCC4}
\end{eqnarray}
the quenching functions $\phi^{v}(\bar{B})$ and $\phi^m(\bar{B})$
are given by
\begin{eqnarray}
\phi^v(\bar{B}) &=& {12 \over 7 \beta^{2}} \, \biggl[1 - {\arctan
\beta \over \beta} \biggr] + {3 \over 7} \tilde L(\beta) \;,
\label{X22} \\
\phi^m(\bar{B}) &=& {3 \over \beta^{2}} \, \biggl[1 - {\arctan
\beta \over \beta} \biggr] \;
\label{X23}
\end{eqnarray}
(see \cite{RK2000}), where $\beta = \sqrt{8} \bar{B} $ and $
\tilde L(y) = 1 - 2 y^{2} + 2 y^{4} \ln (1 + y^{-2}) .$ Thus
$\phi^v(\bar{B}) = 2/\beta^2$ and $\phi^m(\bar{B}) = 3/\beta^2$
for $\beta \gg 1 ;$ and $\phi^v(\bar{B}) = 1-(6/5)\beta^2$ and
$\phi^m(\bar{B}) = 1-(3/5)\beta^2$ for $\beta \ll 1 .$ The
quenching functions $\phi^\Omega_1(\bar{B})$ and
$\phi^\Omega_2(\bar{B})$ are determined by Eqs.~(\ref{SS20}) and
~(\ref{SS21}) in Appendix A. The function $\chi^c(\bar{\bf B})$
entering the magnetic part of the $\alpha$ effect is determined by
the dynamical equation~(\ref{D1}). Note that in Eq.~(\ref{C4}) we
neglected small contributions $\sim O(\delta \Omega / \Omega)$
caused by the mean differential rotation and inhomogeneity of
turbulence [these effects are given by Eqs.~(\ref{K8})-(\ref{K10})
in Appendix A]. For a nonhelical background turbulence the first
term, $\chi^v \phi^v(\bar{B})$, in Eq.~(\ref{C4}) vanishes.

The contribution to the nonlinear $\alpha$ effect caused by a
uniform rotation for a weak mean magnetic field $\bar{B} \ll
\bar{B}_{\rm eq} / 4$ is given by
\begin{eqnarray}
\alpha^\Omega &=& - {16 \over 15} \, {L \, \Omega_\ast \over
L_{_{T}}} \, \bec{\hat{\omega}} {\bf \cdot} \biggl[\, {\bf
\Lambda}^{(v)} - {\epsilon \over 3} \, {\bf \Lambda}^{(b)}
\nonumber\\
&& - {180 \over 7} \, \biggl({\bf \Lambda}^{(v)} - {3 \epsilon
\over 5} \, {\bf \Lambda}^{(b)} \biggr) \, \bar{B}^2 \biggr] \;,
\label{X15}
\end{eqnarray}
and $\bar{B} \gg \bar{B}_{\rm eq} / 4$ it is given by
\begin{eqnarray}
\alpha^\Omega &=& - {1 \over 3 \beta^2} \, {L \, \Omega_\ast \over
L_{_{T}}} \, \bec{\hat{\omega}} {\bf \cdot} ({\bf \Lambda}^{(v)} +
\epsilon \, {\bf \Lambda}^{(b)})
\nonumber\\
&& - {11 \, \Omega_\ast \over 3 \beta} \, (\bec{\hat{\omega}} {\bf
\cdot} {\bf \Lambda}^{(B)}) \, (1 - 1.3 \epsilon) \;, \label{X16}
\end{eqnarray}
[see Eqs.~(\ref{SS12}) and~(\ref{SS15}) in Appendix A], where the
parameter $\epsilon = \langle {\bf b}^2 \rangle^{(0)} / \langle
{\bf u}^2 \rangle^{(0)}$ is the ratio of the magnetic and kinetic
energies in the background turbulence. Asymptotic formula
(\ref{X15}) for $\alpha^\Omega$ in the limit of a very small mean
magnetic field coincides with that obtained in \cite{RKR03} for
$q=5/3$.

The splitting of the nonlinear $\alpha$ effect into the
hydrodynamic, $\alpha^v $, and magnetic, $\alpha^m $, parts was
first suggested in \cite{PFL76}. The magnetic part $\alpha^m$
includes two types of nonlinearity: the algebraic quenching
described by the function $\phi^m(\bar{B})$ (see
\cite{FB99,RK2000}) and the dynamic nonlinearity which is
determined by Eq.~(\ref{D1}). The algebraic quenching of the
$\alpha$-effect is caused by the direct and indirect modification
of the electromotive force by the mean magnetic field. The
indirect modification of the electromotive force is caused by the
effect of the mean magnetic field on the velocity fluctuations and
on the magnetic fluctuations, while the direct modification is due
to the effect of the mean magnetic field on the cross-helicity
(see \cite{RK2000,RK2001}).

The nonlinear turbulent magnetic diffusion coefficients of the
mean magnetic field are given by
\begin{eqnarray}
\eta_{_{A}}(\bar{B}) &=& \phi_1(\bar{B}) \;, \quad
\eta_{_{B}}(\bar{B}) = \phi_1(\bar{B}) + \phi_3(\bar{B}) \;
\label{L3}
\end{eqnarray}
(see \cite{RK04}), where the quenching functions $\phi_k(\bar{B})$
are given by
\begin{eqnarray}
\phi_1(\bar{B}) &=& A_{1}^{(1)}(4 \bar{B}) + A_{2}^{(1)}(4
\bar{B}) \;,
\label{LLL3}\\
\phi_2(\bar{B}) &=& - {1 \over 2} (1 + \epsilon) A_{2}^{(1)}(4
\bar{B}) \;,
\nonumber\\
\phi_3(\bar{B}) &=& (2 - 3 \epsilon) A_{2}^{(1)}(4 \bar B) - (1 -
\epsilon) {3 \over 2 \pi} \bar A_{2}(16 \bar B^2) \;, \nonumber
\end{eqnarray}
the functions $\bar A_{k}(y)$ and $A_{k}^{(1)}(y)$ are given by
Eqs.~(\ref{X24})-(\ref{X27}) in Appendix A. The asymptotic
formulas for the functions $\phi_k(\bar{B})$ for $\bar{B} \ll
\bar{B}_{\rm eq} / 4$ are given by $\phi_1(\bar{B}) = 1 - (12 / 5)
\, \beta^{2} $, $\, \phi_2(\bar{B}) = 1 - (4 / 5) \, (1 +
\epsilon) \, \beta^{2}$ and $\phi_3(\bar{B}) = - (8 / 5) \, (1 - 2
\epsilon) \, \beta^{2} $. For $\bar{B} \gg \bar{B}_{\rm eq} / 4$
they are given by $\phi_1(\bar{B}) =  1 / \beta^2$, $\,
\phi_2(\bar{B}) = \phi_3(\bar{B}) = 2 (1 + \epsilon) / 3 \beta$,
where $\beta = \sqrt{8} \bar{B}$. Note that in Eq.~(\ref{L3}) we
neglected small contributions $\sim O[(l_0 / L)^2]$ caused by the
mean differential rotation.

\subsection{The nonlinear coefficients $\sigma_0(\bar{B})$
and $\sigma_1(\bar{B})$ defining the shear-current effect and the
nonhelical $\alpha$ effect}

The nonlinear coefficient $\sigma_0(\bar{B})$ describes the
shear-current effect (see \cite{RK03,RK04}) and
$\sigma_1(\bar{B})$ determines the nonhelical $\alpha$ effect. The
parameters $\sigma_0(\bar{B})$ and $\sigma_1(\bar{B})$ are
determined by the corresponding contributions from the
$\bec{\delta}(\bar{\bf B})$ term, the ${\eta}_{ij}(\bar{\bf B})$
term and the ${\kappa}_{ijk}(\bar{\bf B})$ term in the nonlinear
electromotive force~(\ref{A14}) caused by the mean differential
rotation. We found that the nonlinear function $\sigma_0(\bar{B})$
defining the shear-current effect is the same for a turbulence
with a mean differential rotation in cylindrical and spherical
geometries for an axisymmetric mean field dynamo problem and for a
nonrotating turbulence with an imposed linear mean velocity shear
in a plane geometry. The latter case was studied in \cite{RK04}.

To explain the physics of the shear-current effect, we compare the
$\alpha$ effect in the $\alpha {\bf \Omega} $ dynamo with the
$\bec{\delta}$ term caused by the shear-current effect (see
\cite{RK03,RK04}). The $\alpha$ term  in the nonlinear
electromotive force which is responsible for the generation of the
mean magnetic field and caused by a uniform rotation and
inhomogeneity of turbulence, reads $ {\cal E}^\alpha_i \equiv
\alpha^v \bar B_i \propto - ({\bf \Omega} \cdot {\bf
\Lambda}^{(v)}) \bar B_i $ (see \cite{RKR03}), where ${\bf
\Lambda}^{(v)}$ determines the inhomogeneity of turbulence. The
$\bec{\delta}$ term in the electromotive force caused by the
shear-current effect is given by $ {\cal E}^\delta_i \equiv -
(\bec{\delta} {\bf \times} (\bec{\nabla} {\bf \times} \bar{\bf
B}))_i \propto - (\bar{\bf W} \cdot \bec{\nabla}) \bar B_i $ (see
\cite{RK03}), where the $\bec{\delta}$ term is proportional to the
mean vorticity $\bar{\bf W} = \bec{\nabla} {\bf \times} \bar {\bf
U}$ which is caused by the differential rotation.

During the generation of the mean magnetic field in both cases (in
the $\alpha {\bf \Omega} $ dynamo and in the shear-current
dynamo), the mean electric current along the original mean
magnetic field arises. The $\alpha$ effect is related to the
hydrodynamic helicity $ \propto ({\bf \Omega} \cdot {\bf
\Lambda}^{(v)}) $ in an inhomogeneous turbulence. The deformations
of the magnetic field lines are caused by upward and downward
rotating turbulent eddies in the $\alpha {\bf \Omega} $ dynamo.
Since the turbulence is inhomogeneous (which breaks a symmetry
between the upward and downward eddies), their total effect on the
mean magnetic field does not vanish and it creates the mean
electric current along the original mean magnetic field (see
\cite{P79}).

In a turbulent flow with the mean differential rotation, the
inhomogeneity of the original mean magnetic field breaks a
symmetry between the influence of upward and downward turbulent
eddies on the mean magnetic field. The deformations of the
magnetic field lines in the shear-current dynamo are caused by
upward and downward turbulent eddies which result in the mean
electric current along the mean magnetic field and produce the
magnetic dynamo (see \cite{RK03,RK04}).

Note that the differential rotation is described by the gradient
tensor of the mean velocity field $\nabla_i \bar U_{j} = (\partial
\hat U)_{ij} + \varepsilon_{ijn} (\bec{\bf \nabla} {\bf \times}
\bar{\bf W})_{n} / 2 $, where the symmetric part of the gradient
tensor $({\partial \hat U})_{ij} = (\nabla_i \bar U_{j} + \nabla_j
\bar U_{i}) / 2$ is given by
\begin{eqnarray}
(\partial \bar U)_{ij} = {1 \over 2} [({\bf e}_z {\bf \times} {\bf
r})_i \nabla_j + ({\bf e}_z {\bf \times} {\bf r})_j \nabla_i] \,
(\delta \Omega) \;,
\label{S60}
\end{eqnarray}
and the mean vorticity $\bar{\bf W}$ in cylindrical coordinates is
given by
\begin{eqnarray}
\bar {\bf W} = - \rho \, ({\bf e}_\rho \, \nabla_z - {\bf e}_z \,
\nabla_\rho) \, (\delta \Omega) \;,
\label{S61}
\end{eqnarray}
and in spherical coordinates the mean vorticity is
\begin{eqnarray}
\bar {\bf W} = r \sin \theta \, ({\bf e}_r \, \nabla_\theta - {\bf
e}_\theta \, \nabla_r) \, (\delta \Omega) \; .
\label{S62}
\end{eqnarray}

The nonlinear coefficients $\sigma_0(\bar{B})$ and
$\sigma_1(\bar{B})$ defining the shear-current effect and the
nonhelical $\alpha$ effect are determined by Eqs.~(\ref{F1})
and~(\ref{F5}) in Appendix A. The nonlinear dependencies of the
parameters $\sigma_0(\bar{B})$ and $\sigma_1(\bar{B})$ are shown
in FIG.~1 for different values of the parameter $\epsilon$. The
background magnetic fluctuations caused by the small-scale dynamo
and described by the parameter $\epsilon$, increase the parameter
$\sigma_0(\bar{B})$. For a weak mean magnetic field $\bar{B} \ll
\bar{B}_{\rm eq} / 4$ the parameter $\sigma_0(\bar{B})$ is given
by $\sigma_0(\bar{B}) = (4 / 45) \, (2 - q + 3 \epsilon)$ (see
\cite{RK04}), where $q$ is the exponent of the energy spectrum of
the background turbulence. The latter equation is in agreement
with that obtained in \cite{RK03} where the case a weak mean
magnetic field and $\epsilon=0$ was considered. In this equation
we neglected small contribution $\sim O[(4 \bar{B} / \bar{B}_{\rm
eq})^2]$. The mean magnetic field is generated due to the
shear-current effect, when $\sigma_0(\bar{B})>$, i.e., when the
exponent of the energy spectrum $q < 2 + 3 \epsilon$. Note that
the parameter $q$ varies in the range $1 < q < 3$. Therefore, when
the level of the background magnetic fluctuations caused by the
small-scale dynamo is larger than $1/3$ of the kinetic energy of
the velocity fluctuations, the mean magnetic field can be
generated due to the shear-current effect for an arbitrary
exponent $q$ of the energy spectrum of the velocity fluctuations
(see \cite{RK04}).

\medskip
\begin{figure}
\centering
\includegraphics[width=8cm]{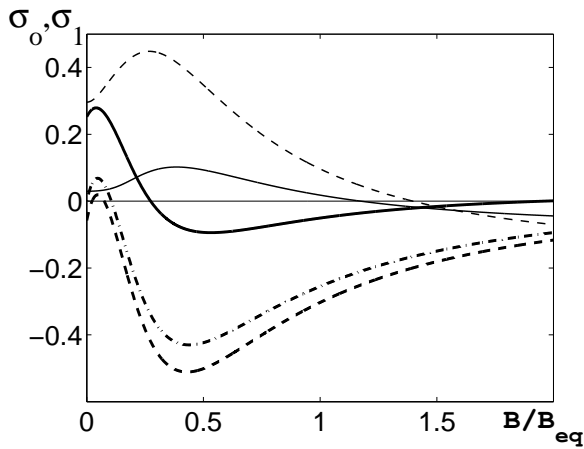}
\caption{\label{Fig1} The nonlinear coefficient
$\sigma_0(\bar{B})$ defining the shear-current effect for
$\epsilon=0$ (thin solid) and for $\epsilon=1$ (thin dashed); and
the nonlinear coefficient $\sigma_1(\bar{B})$ defining the
nonhelical $\alpha$ effect for different values of the parameter
$\epsilon$: $\, \, \, \epsilon=0$ (thick solid); $\epsilon = 17/21
$ (thick dashed-dotted); $\epsilon=1$ (thick dashed).}
\end{figure}

For the Kolmogorov turbulence, i.e., when the exponent of the
energy spectrum of the background turbulence $q=5/3$, the
parameters $\sigma_0(\bar{B})$  and $\sigma_1(\bar{B})$ for
$\bar{B} \ll \bar{B}_{\rm eq} / 4$ are given by $\sigma_0(\bar{B})
= (4 / 135) \, (1 + 9 \epsilon) $ and $ \sigma_1(\bar{B}) = (2 /
135) \, (17 - 21 \epsilon)$. For $\bar{B} \gg \bar{B}_{\rm eq} /
4$ they are given by $\sigma_0(\bar{B}) = - (11 / 135) \, (1 +
\epsilon)$ and $\sigma_1(\bar{B}) = (2 / 135) \, (1 + \epsilon)$.
It is seen from these equations and from FIG. 1 that the nonlinear
coefficient $\sigma_0(\bar{B})$ changes its sign at some value of
the mean magnetic field $\bar{B}=\bar{B}_\ast$. For instance,
$\bar{B}_\ast = 0.6 \bar{B}_{\rm eq}$ for $\epsilon=0$, and
$\bar{B}_\ast = 0.3 \bar{B}_{\rm eq}$ for $\epsilon=1$. However,
there is no quenching of this effect contrary to the quenching of
the nonlinear $\alpha$ effect, the nonlinear turbulent magnetic
diffusion, the nonlinear ${\bf \Omega} {\bf \times} \bar{\bf J}$
effect, etc.

The mean differential rotation causes the nonhelical $\alpha$
effect, $W_\ast \, \sigma_1(\bar{\bf B}) \, \nabla_z (\delta
\Omega)$ [see Eqs.~(\ref{L8}) and~(\ref{L10})], which is
independent of a hydrodynamic helicity. It follows from the
asymptotic formula for $\sigma_1(\bar{B})$ at $\bar{B} \gg
\bar{B}_{\rm eq} / 4$ that there is no quenching of this effect
contrary to the quenching of the regular nonlinear $\alpha$ effect
(see Section III-A). These two kinds of the $\alpha$ effect have
opposite signs. Thus, the total $\alpha$ effect should change its
sign during the nonlinear growth of the mean magnetic field. The
nonhelical $\alpha$ effect vanishes if the mean rotation is
constant on the cylinders which are parallel to the rotation axis.
Note that $\sigma_1(\bar B=0.1 \bar{B}_{\rm eq}) = 0$ for
$\epsilon=1$.

The $\bec{\delta}$ term in the electromotive force which is
responsible for the shear-current effect has been also calculated in
\cite{RS05,RKICH05} for a kinematic problem using the second-order
correlation approximation (SOCA). However, these studies did not
found the dynamo action in nonrotating and nonhelical shear flows.
Note that the second order correlation approximation (SOCA) is valid
for small hydrodynamic Reynolds numbers. Indeed, even in a highly
conductivity limit (large magnetic Reynolds numbers) SOCA can be
valid only for small Strouhal numbers, while for large hydrodynamic
Reynolds numbers (fully developed turbulence) the Strouhal number is
unity. Our recent studies for small hydrodynamic and magnetic
Reynolds numbers (using spectral $\tau$ approximation) also did not
found the dynamo action in nonrotating and nonhelical shear flows in
agreement with \cite{RS05,RKICH05}.

\subsection{The nonlinear coefficient $\delta_0^\Omega(\bar{B})$
defining the ${\bf \Omega} {\bf \times} \bar{\bf J}$ effect}

\begin{figure}
\centering
\includegraphics[width=8cm]{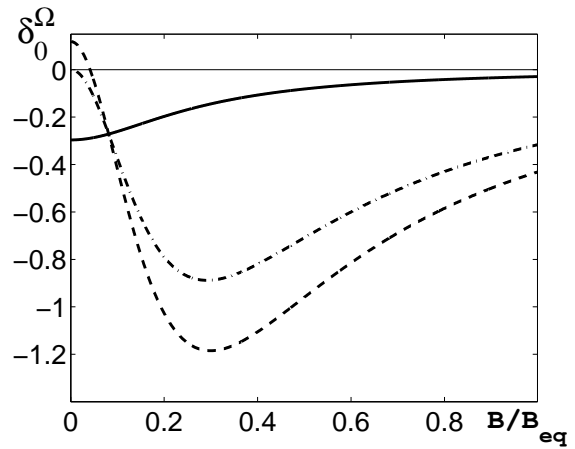}
\caption{\label{Fig2} The nonlinear coefficient
$\delta_0^\Omega(\bar{B})$ defining the ${\bf \Omega} {\bf \times}
\bar{\bf J}$ effect for different values of the parameter
$\epsilon$: $\, \, \, \epsilon=0$ (dashed); $\epsilon = 2/7$
(dashed-dotted); $\epsilon=1$ (solid).}
\end{figure}

The $\bec{\delta}$ term in the electromotive force which is caused
by a uniform rotation, describes the ${\bf \Omega \times \bar J}$
effect. This effect in combination with the differential rotation
can cause a generation of the mean magnetic field even in a
nonhelical turbulent flow (see \cite{R69,R72,MP82,R86} and
\cite{RKR03}), where $\bar{\bf J}$ is the mean electric current.
The nonlinear coefficient $\delta_0^\Omega(\bar{B})$ defining the
${\bf \Omega} {\bf \times} \bar{\bf J}$ effect is determined by
\begin{eqnarray}
\delta_0^\Omega(\bar{B}) &=& - {2 \over 3} \biggl[\Psi_4\{C_{1}+
C_{3}\}_y - (1 - \epsilon) (\Psi_2 + 4 \Psi_3)\{A_{1}
\nonumber \\
&& + A_{2}\}_y + (1 + \epsilon) (A_{1}^{(2)}(y) + A_{2}^{(2)}(y))
\biggr]_{y=4\bar{B}} \;,
\nonumber \\
\label{M1}
\end{eqnarray}
where the functions $\Psi_k\{X\}_y$ are determined by
Eqs.~(\ref{X1}) in Appendix A. The parameter
$\delta_0^\Omega(\bar{B})$ is determined by the contributions from
the $\bec{\delta}(\bar{\bf B})$ term, the ${\eta}_{ij}(\bar{\bf
B})$ term and the ${\kappa}_{ijk}(\bar{\bf B})$ term in the
nonlinear electromotive force~(\ref{A14}) caused by a uniform
rotation. The nonlinear coefficient $\delta_0^\Omega(\bar{B})$ is
shown in FIG. 2 for different values of the parameter $\epsilon$.
The asymptotic formulas for the coefficient
$\delta_0^\Omega(\bar{B})$ for a weak mean magnetic field $\bar{B}
\ll \bar{B}_{\rm eq} / 4$ are
\begin{eqnarray}
\delta_0^\Omega(\bar{B}) &=& {8 \over 135} (2 - 7 \epsilon) \;,
\label{MM2}
\end{eqnarray}
and for $\bar{B} \gg \bar{B}_{\rm eq} / 4$ are
\begin{eqnarray}
\delta_0^\Omega(\bar{B}) &=& - {1 \over 3 \beta^2} (34 +19
\epsilon) \; . \label{M3}
\end{eqnarray}
Asymptotic formula~(\ref{MM2}) for a weak mean magnetic field
$(\bar{B} \ll \bar{B}_{\rm eq} / 4)$ coincide with that obtained
in \cite{RKR03} for $q=5/3$.

\subsection{The dynamical equation for the function
$\chi^c(\bar{\bf B})$}

The function $\chi^c(\bar{\bf B})$ entering the magnetic part of
the $\alpha$ effect [see Eq.~(\ref{C4})] is determined by the
dynamical equation
\begin{eqnarray}
{\partial \chi^{c} \over \partial t} &=& - 4 \biggl({L \over l_0}
\biggr)^2 [\bec{\cal E} {\bf \cdot} \bar{\bf B} + \bec{\nabla}
\cdot \bec{\cal F}^{(\chi)}]
\nonumber \\
& & - \bec{\nabla} \cdot (\bar{\bf U} \chi^{c}) - \chi^{c} / T \;
, \label{D1}
\end{eqnarray}
(see, e.g., \cite{KR82,KR99}), where $\bec{\cal F}^{(\chi)}$ is
the nonadvective flux of the magnetic helicity which serves as an
additional nonlinear source in the equation for $ \chi^{c} $ (see
\cite{KMRS2000,KMRS02}), $ \, \bar{\bf U} \chi^{c}$ is the
advective flux of the magnetic helicity, $\bar{\bf U}$ is the
differential rotation, and $ T = (1/3) (l_0/L)^{2} {\rm Rm} $ is
the characteristic time of relaxation of magnetic helicity.
Equation~(\ref{D1}) was obtained using arguments based on the
magnetic helicity conservation law. The function $\chi^{c}$ is
proportional to the magnetic helicity, $\chi^{c} = 2 \chi^{m} / (9
\mu \eta_{_{T}} \rho_0)$ (see \cite{KR99}), where $\chi^{m} =
\langle {\bf a} \cdot {\bf b} \rangle$ is the magnetic helicity
and ${\bf a}$ is the vector potential of small-scale magnetic
field. The physical meaning of Eq.~(\ref{D1}) is that the total
magnetic helicity is a conserved quantity and if the large-scale
magnetic helicity grows with mean magnetic field, the evolution of
the small-scale helicity should somehow compensate this growth.
Compensation mechanisms include dissipation and various kinds of
transport (see \cite{KMRS2000,KMRS02}).

In order to demonstrate an important role of the nonadvective flux
of the magnetic helicity, let us consider a local model in
cylindrical coordinates, when the mean magnetic field depend only
on the vertical coordinate $z$ and $A' \gg A / r$, where $A' =
\partial A / \partial z$ and $\bar{\bf B} = B {\bf e}_\varphi
- A' {\bf e}_r $. Since
\begin{eqnarray}
{\partial A \over \partial t} = {\cal E}_\varphi \;,
\label{D3}\\
{\partial B \over \partial t} = {\cal E}'_r - D A' \;, \label{D4}
\end{eqnarray}
we obtain that
\begin{eqnarray}
\bec{\cal E} {\bf \cdot} \bar{\bf B} = {\cal E}_\varphi B - {\cal
E}_r A' = B {\partial A \over \partial t} + {\cal E}_r {1 \over D}
{\partial B \over \partial t}  - {1 \over 2 D} ({\cal E}_r^2)' \;,
\nonumber \\
\label{DD5}
\end{eqnarray}
Then in steady state Eqs.~(\ref{D1}) and (\ref{DD5}) yield
\begin{eqnarray}
{\cal E}_r^2 - 2 D {\cal F}^{(\chi)}_z = const \;, \label{D6}
\end{eqnarray}
where ${\cal E}_r = \eta_{_{B}} B'$. Here we neglected the last
term in Eq.~(\ref{D1}) which, e.g., for galactic dynamo is very
small. In a steady-state for fields of even parity with respect to
the disc plane, we obtain the solution of Eq.~(\ref{D6}) for
positive $ C \, D $
\begin{eqnarray}
\int_{0}^{B} {\eta_{_{B}}(\tilde B) \over \sqrt{|{\cal F}(\tilde
B)}|} \,d \tilde B = \sqrt{2 |C \, D|} \int_{|z|}^{1}
\sqrt{|\chi^v(\tilde z)|} \,d \tilde z \;, \label{C2}
\end{eqnarray}
where ${\cal F}^{(\chi)}_z = C |{\cal F}(B)| |\chi^v(z)| .$ The
crucial point for the dynamo saturation is a nonzero flux of
magnetic helicity. It follows from Eq.~(\ref{C2}) that this
saturation mechanism is nearly independent of the form of the flux
of magnetic helicity. In that sense this is a universal mechanism
which limits growth of the mean magnetic field. If we assume that
$|\bec{\cal F}(\bar B)| \sim \bar B^{-2 \gamma_\ast}$, we obtain
that the saturated mean magnetic field is
\begin{eqnarray}
\bar B_\varphi = |C \, D|^{1 \over 2\gamma_\ast} \biggl[
\int_{|z|}^{1} \sqrt{|\chi^v(\tilde z)|} \,d \tilde z \biggr]^{1
\over \gamma_\ast} \, \bar B_{\rm eq} \;, \label{AA8}
\end{eqnarray}
where we redefined the constant $C$, we took into account that
$\eta_{_{B}}(\bar B) \propto B_{\rm eq} / \bar B$ for $\bar B \gg
\bar B_{\rm eq} / 4 $, and we restored the dimensional factor
$\bar B_{\rm eq} .$ Note that the nonadvective flux of the
magnetic helicity was chosen in \cite{KMRS02} in the form $
\bec{\cal F}^{(\chi)} = C \chi^v \phi_{v}(\bar{B}) \bar{\bf B}^2
\eta_{A}^{(z)}(\bar{B}) (\bec{\nabla} \rho_0) / \rho_0 $. This
corresponds to $\gamma_\ast = 1$ in the function $|\bec{\cal
F}(B)| $. For the specific choice of the profile $|\chi^v(z)|
=\sin^{2}(\pi z/ 2)$ we obtain
\begin{eqnarray}
\bar B_\varphi &\approx & {4 \over 1 + \epsilon} \sqrt{|C \, D|}
\, \bar B_{\rm eq} \cos \, \biggl({{\pi z} \over 2} \biggr) \;,
\label{C6} \\
\bar B_r &\approx & - {1 + \epsilon \over 4 |R_\omega|} \bar
B_{\rm eq} \tan \, \biggl({{\pi z} \over 2} \biggr) \;, \label{C7}
\end{eqnarray}
where we have now restored the dimensional factor  $\bar B_{\rm
eq} .$ The boundary conditions for $\bar B_\varphi$ are $\bar
B_\varphi (z=1) = 0 $, $\, B_\varphi'(z=0) = 0 $, and for $B_{r}$
are $B_{r}(z=1) = 0$, $\, B_{r}'(z=0) = 0 .$ Note, however, that
this asymptotic analysis performed for $\bar B \gg \bar B_{\rm eq}
/ 4 $  is not valid in the vicinity of the point $z=1$ because
$\bar B(z=1) = 0$.

\subsection{The dynamo waves}

In order to elucidate the new effects caused by the differential
rotation, let us consider first a kinematic problem in a spherical
geometry. Following \cite{P55} we study dynamo action in a thin
convective shell, average the linearized equations~(\ref{L10})
and~(\ref{L11}) for $A$ and $B$ over the depth of the convective
shell. Then we neglect the curvature of the convective shell and
replace it by a flat slab. These equations are obviously
oversimplified. However, they can be used to reproduce basic
qualitative features of solar and stellar activity (see, e.g.,
\cite{KKMR03}). We are interested in dynamo waves propagating from
middle solar latitudes towards the equator. We seek for a solution
of the obtained equations in the form of the growing waves, $A,B
\propto \exp (\gamma \, t) \exp[i(\omega \, t - {\bf K} \cdot {\bf
R})]$, where the growth rate of the dynamo waves with the
frequency
\begin{eqnarray}
\omega = - \alpha_l \, \sqrt{{D \, |S_K| \over 2}} \, {{\rm
sgn}(S_K) \over \sqrt{\sigma_l + \sqrt{\sigma_l^2 + \alpha_l^2}}}
\label{R20}
\end{eqnarray}
is given by
\begin{eqnarray}
\gamma = \sqrt{{D \, |S_K| \over 2} \, \biggl[\sigma_l +
\sqrt{\sigma_l^2 + \alpha_l^2} \biggr]} - K^2 \; . \label{R21}
\end{eqnarray}
The frequency and the growth rate of the dynamo waves are written
in a dimensionless form. Here
\begin{eqnarray*}
\alpha_l &=& \alpha + W_\ast \, \sigma_1 \, S_z \;,
\\
\sigma_l &=& W_\ast \, \sigma_0 \, S_K + \Omega_\ast \,
\delta_0^\Omega \, K_z \;,
\\
S_K &=& K_\theta \, \nabla_r (\delta \Omega)  - K_r \,
\nabla_\theta (\delta \Omega) \;,
\\
S_z  &=& \cos \theta \, \nabla_r (\delta \Omega)  - \sin \theta \,
\nabla_\theta (\delta \Omega) \;,
\\
K_z  &=& \cos \theta \, K_r - \sin \theta \, K_\theta \; .
\end{eqnarray*}
The total $\alpha$ effect, $\alpha_l$, is a sum of the usual
$\alpha$ effect (caused by helical motions) and a nonhelical
contribution, $W_\ast \, \sigma_1 \, S_z$, due to the effect of
the the mean differential rotation on the small-scale turbulence.
The parameter $\sigma_l$ describes both, the shear-current effect
determined by $W_\ast \, \sigma_0 \, S_K$ term, and the ${\bf
\Omega} {\bf \times} \bar{\bf J}$ effect determined by
$\Omega_\ast \, \delta_0^\Omega \, K_z$ term. Even in nonhelical
turbulent motions, the mean magnetic field is generated due to the
shear-current effect and the ${\bf \Omega} {\bf \times} \bar{\bf
J}$ effect.

\section{DISCUSSION}

Let us discuss the nonlinear effects. It was shown recently in
\cite{RK2001} that the algebraic nonlinearity alone (i.e.,
algebraic quenching of both, the $\alpha$ effect and turbulent
magnetic diffusion) cannot saturate the growth of the mean
magnetic field. Note that the saturation of the growth of the mean
magnetic field in the case with only an algebraic nonlinearity
present can be achieved when the derivative of the nonlinear
dynamo number with respect to the mean magnetic field is negative,
i.e., $dD_N(\bar{B}) / d\bar{B} < 0$.  Here $D_N(\bar{B}) =
\alpha(\bar{B}) / [\eta_{_{A}}(\bar{B}) \, \eta_{_{B}}(\bar{B})] $
is the nonlinear dynamo number. Thus, when the nonlinear dynamo
number decreases with the growth of the mean magnetic field, the
nonlinear saturation of the magnetic field is possible.

In this study we showed that the differential rotation of fluid
can decrease the total $\alpha$ effect. In particular, the mean
differential rotation causes the nonhelical $\alpha$ effect,
$W_\ast \, \sigma_1(\bar{\bf B}) \, \nabla_z (\delta \Omega)$,
which is independent of a hydrodynamic helicity. We demonstrated
that there is no quenching of this effect contrary to the
quenching of the regular nonlinear $\alpha$ effect,
$\alpha(\bar{\bf B}) = \chi^v \phi^v(\bar{B}) + \alpha^\Omega +
\chi^c(\bar{\bf B}) \phi^m(\bar{B})$. In this study we found that
these two kinds of the $\alpha$ effect have opposite signs. Thus,
the total $\alpha$ effect should change its sign during the
nonlinear evolution of the mean magnetic field, and there is a
range of magnitudes of the mean magnetic field, where the
nonlinear dynamo number decreases with the growth of the mean
magnetic field. Therefore, the algebraic nonlinearity alone can
saturate the growth of the mean magnetic field if one take into
account the effect of differential rotation on the nonlinear
electromotive force. For instance, the nonhelical $\alpha$ effect
causes a saturation of the growth of the mean magnetic field at
the base of the convective zone at $ \bar B \leq 2 \bar B_{\rm
eq}$ (see below), where $\bar B_{\rm eq}$ is the equipartition
mean magnetic field. However, the nonhelical $\alpha$ effect
vanishes if the mean rotation is constant on the cylinders which
are parallel to the rotation axis.

In this study we also demonstrated that the mean differential
rotation which causes the shear-current effect, increases a growth
rate of the large-scale dynamo instability at weak mean magnetic
fields, and causes a saturation of the growth of the mean magnetic
field for a stronger field.

The nonlinear shear-current effect and the nonhelical $\alpha$
effect become very important at the base of the convective zone
(see below). When we apply the obtained results to the solar
convective zone, we have to take into account that all physical
ingredients of the dynamo model vary strongly with the depth $H$
below the solar surface and we have to use some average quantities
in the dynamo equations. We use mainly estimates of governing
parameters taken from models of the solar convective zone (see,
e.g., \cite{S74,BT66}). In particular, in the upper part of the
convective zone, say at depth $ H \sim 2 \times 10^7$ cm, the
magnetic Reynolds number $ {\rm Rm} \sim 10^5 ,$ the maximum scale
of turbulent motions $ l_0 \sim 2.6 \times 10^7$ cm, the
characteristic turbulent velocity in the maximum scale $l_{0}$ of
turbulent motions $u_0 \sim 9.4 \times 10^4 $ cm s$^{-1}$, the
fluid density $\rho_0 \sim 4.5 \times 10^{-7}$ g cm$^{-3} ,$ the
turbulent magnetic diffusion $\eta_{_{T}} \sim 0.8 \times 10^{12}
$ cm$^2$ s$^{-1}$ and the equipartition mean magnetic field is
$\bar{B}_{\rm eq} = 220 $ G. Thus, in the upper part of the
convective zone the parameters $W_\ast \sim 10^{-3} - 10^{-4}$ and
$\Omega_\ast \sim 5 \times (10^{-3} - 10^{-4}) $. According to
various models, the ranges of the dynamo number $D \approx 10^3 -
10^6$ can be considered as realistic for the solar case. At the
base of the convective zone (at depth $ H \sim 2 \times 10^{10}$
cm), the magnetic Reynolds number ${\rm Rm} = l_0 u_0 / \eta \sim
2 \cdot 10^9 ,$ the maximum scale of turbulent motions $ l_0 \sim
8 \times 10^9$ cm, the characteristic turbulent velocity $u_0 \sim
2 \times 10^3 $ cm s$^{-1}$, the fluid density $ \rho_0 \sim 2
\times 10^{-1}$ g cm$^{-3} ,$ the turbulent magnetic diffusion
$\eta_{_{T}} \sim 5.3 \times 10^{12} $ cm$^2$s$^{-1}$. The
equipartition mean magnetic field $\bar{B}_{\rm eq} = 3000 $ G.
Thus, at the base of the convective zone the parameters $W_\ast
\sim 1 - 10 $ and $\Omega_\ast \sim 5 - 50$. Thus, the effects of
the differential rotation (the nonlinear shear-current effect and
the nonhelical $\alpha$ effect) become very important at the base
of the convective zone. Since these effects are not quenched, they
might be the only surviving effects.

\appendix

\section{Effects of uniform and differential rotations}

The method of the derivation of equation for the nonlinear
electromotive force in a rotating turbulence is similar to that
used in \cite{RK04} for a nonrotating turbulence with an imposed
mean velocity shear. In the framework of a mean-field approach we
derive equations for the following correlation functions:
$f_{ij}({\bf k}) = \hat L(u_i; u_j)$, $h_{ij}({\bf k}) = \hat
L(b_i; b_j)$ and $g_{ij}({\bf k}) = \hat L(b_i; u_j)$, where $\hat
L(a_i; c_j)$ is determined by Eq.~(\ref{X20}). In order to exclude
the pressure term from the equation of motion~(\ref{B1}) we
calculate $ \bec{\nabla} {\bf \times} (\bec{\nabla} {\bf \times}
{\bf u}) .$ Then we rewrite the obtained equation and
Eq.~(\ref{B2}) in a Fourier space. The equations for these
correlation functions are given by
\begin{eqnarray}
{\partial f_{ij}({\bf k}) \over \partial t} &=& M_{ijpq}^\Omega
f_{pq} + I_{ijmn}^\sigma(\bar{\bf U}) f_{mn}
\nonumber \\
&& + i({\bf k} {\bf \cdot} \bar{\bf B}) \Phi_{ij}^{(M)} + I^f_{ij}
+ F_{ij} + \hat D f_{ij}^N \;,
\label{B6} \\
{\partial h_{ij}({\bf k}) \over \partial t} &=&
E_{ijmn}^\sigma(\bar{\bf U}) h_{mn} - i({\bf k}{\bf \cdot}
\bar{\bf B}) \Phi_{ij}^{(M)}
\nonumber \\
&& + I^h_{ij} + \hat D h_{ij}^N \;,
\label{B7} \\
{\partial g_{ij}({\bf k }) \over \partial t} &=& D_{jn}^\Omega
g_{in} + J_{ijmn}^\sigma(\bar{\bf U}) g_{mn} + i({\bf k} {\bf
\cdot} \bar{\bf B}) [f_{ij}({\bf k})
\nonumber \\
&& - h_{ij}({\bf k}) - h_{ij}^{(H)}] + I^g_{ij} + \hat D g_{ij}^N
\;, \label{B8}
\end{eqnarray}
where the mean velocity $ \bar {\bf U} $ describes the
differential rotation, $\Phi_{ij}^{(M)}({\bf k}) = g_{ij}({\bf k})
- g_{ji}(-{\bf k}) ,$ $ \, F_{ij}({\bf k}) = \langle \tilde F_i
({\bf k}) u_j(-{\bf k}) \rangle + \langle u_i({\bf k}) \tilde
F_j(-{\bf k}) \rangle$, $\, {\bf \tilde F} ({\bf k}) = {\bf k}
{\bf \times} ({\bf k} {\bf \times} {\bf F}({\bf k}))/k^2 \rho_0 .$
The tensors $M_{ijpq}^\Omega$ and $D_{ij}^\Omega$ are given by
\begin{eqnarray*}
M_{ijpq}^\Omega &=& D_{ip}({\bf k}_1) \delta_{jq} + D_{jq}({\bf
k}_2) \delta_{ip} = \hat M_{ijpq}^\Omega + \tilde M_{ijpq}^\Omega
\;,
\\
D_{ij}^\Omega &=& D_{ij}({\bf k}_2) = \hat D_{ij}^\Omega + \tilde
D_{ij}^\Omega  \;,
\\
\hat M_{ijpq}^\Omega &=& 2 \Omega_m k_{mn} (\varepsilon_{ipn}
\delta_{jq} + \varepsilon_{iqn} \delta_{ip}) \;,
\\
\tilde M_{ijpq}^\Omega &=& - 2 i \Omega_m T_{mnl}
(\varepsilon_{ipn} \delta_{jq} - \varepsilon_{iqn} \delta_{ip})
\nabla_l \;,
\\
\hat D_{ij}^\Omega &=& 2 \varepsilon_{ijm} \Omega_m k_{mn} \;,
\quad \tilde D_{ij}^\Omega = 2 i \varepsilon_{ijm} \Omega_m
T_{mnl} \nabla_l \;,
\\
T_{mnp} &=& (1/2 k^2) (k_m \delta_{np} + k_n \delta_{mp} - 2k \,
k_{mnp}) \;,
\end{eqnarray*}
where $D_{ij}({\bf k}) = 2 \varepsilon_{ijm} k_{m} ({\bf k} \cdot
{\bf \Omega}) / k^{2} $. The tensors $I_{ijmn}^\sigma(\bar{\bf
U})$, $\, E_{ijmn}^\sigma(\bar{\bf U})$ and
$J_{ijmn}^\sigma(\bar{\bf U})$ are given by
\begin{eqnarray*}
I_{ijmn}^\sigma(\bar{\bf U}) &=& \biggl[2 k_{iq} \delta_{mp}
\delta_{jn} + 2 k_{jq} \delta_{im} \delta_{pn} - \delta_{im}
\delta_{jq} \delta_{np}
\nonumber\\
&& - \delta_{iq} \delta_{jn} \delta_{mp} + \delta_{im} \delta_{jn}
k_{q} {\partial \over \partial k_{p}} \biggr] \nabla_{p} \bar
U_{q} \;,
\nonumber\\
E_{ijmn}^\sigma(\bar{\bf U}) &=& (\delta_{im} \delta_{jq} +
\delta_{jm} \delta_{iq}) \, \nabla_{n} \bar U_{q} \;,
\nonumber\\
J_{ijmn}^\sigma(\bar{\bf U}) &=& \biggl[2 k_{jq} \delta_{im}
\delta_{pn} - \delta_{im} \delta_{pn} \delta_{jq} + \delta_{jn}
\delta_{pm} \delta_{iq}
\nonumber\\
& & + \delta_{im} \delta_{jn} k_{q} {\partial \over \partial
k_{p}} \biggr] \nabla_{p} \bar U_{q} \;
\end{eqnarray*}
(see \cite{RK03,RK04}), where $\delta_{ij}$ is the Kronecker
tensor, $ k_{ij} = k_i k_j / k^2 $. Equation~(\ref{B6})-(\ref{B8})
are written in a frame moving with a local velocity $ \bar {\bf U}
$. For the derivation of Eqs.~(\ref{B6})-(\ref{B8}) we used the
relation
\begin{eqnarray*}
\varepsilon_{ijn} \Omega_n k^2 + (\varepsilon_{inl} k_j -
\varepsilon_{jnl} k_i ) k_n \Omega_l = \varepsilon_{ijn} k_n ({\bf
k} \cdot {\bf \Omega}) \;,
\end{eqnarray*}
which applies to arbitrary vectors ${\bf k}$ and ${\bf \Omega}$
(see \cite{RKR03}). The source terms $I_{ij}^f$ , $\, I_{ij}^h$
and $I_{ij}^g$ (which contain the large-scale spatial derivatives
of the mean magnetic field and the second moments) are given by
\begin{eqnarray}
I_{ij}^f &=& {1 \over 2}(\bar{\bf B} {\bf \cdot} \bec{\nabla})
\Phi_{ij}^{(P)} + [g_{qj}({\bf k}) (2 P_{in}(k) - \delta_{in})
\nonumber \\
&& + g_{qi}(-{\bf k}) (2 P_{jn}(k) - \delta_{jn})] \bar{B}_{n,q} -
\bar{B}_{n,q} k_{n} \Phi_{ijq}^{(P)}\;,
\nonumber\\
\label{MM1}\\
I_{ij}^h &=& {1 \over 2}(\bar{\bf B} {\bf \cdot} \bec{\nabla})
\Phi_{ij}^{(P)} - [g_{iq}({\bf k}) \delta_{jn} + g_{jq}(-{\bf k})
\delta_{in}] \bar{B}_{n,q}
\nonumber\\
&& - \bar{B}_{n,q} k_{n} \Phi_{ijq}^{(P)} \;,
\label{M2}\\
I_{ij}^g &=& {1 \over 2} (\bar{\bf B} {\bf \cdot} \bec{\nabla})
(f_{ij} + h_{ij}) + h_{iq} (2 P_{jn}(k) - \delta_{jn})
\bar{B}_{n,q}
\nonumber \\
&& - f_{nj} \bar{B}_{i,n} - \bar{B}_{n,q} k_{n}(f_{ijq} + h_{ijq})
\; \label{MM3}
\end{eqnarray}
(see \cite{RK04}), where $ \bec{\nabla} =
\partial / \partial {\bf R} $, $\, \Phi_{ij}^{(P)}({\bf k}) =
g_{ij}({\bf k}) + g_{ji}(-{\bf k}) ,$ and $\bar{B}_{i,j} =
\nabla_j \bar{B}_{i} ,$ $ \, f_{ij}^{N} ,$ $\, h_{ij}^{N} $ and $
g_{ij}^{N} $ are the third moments appearing due to the nonlinear
terms, $f_{ijq} = (1/2) \partial f_{ij} / \partial k_{q} ,$ and
similarly for $h_{ijq}$ and $\Phi_{ijq}^{(P)}$. To derive Eqs.
(\ref{B6})-(\ref{B8}) we used the identity:
\begin{eqnarray}
&& i \int \,d {\bf  K} \,d {\bf Q} \, \, (k_{p} + K_{p}/2) \bar
B_{p}({\bf Q}) \exp(i {\bf K} {\bf \cdot} {\bf R})
\nonumber\\
&& \times \langle u_i ({\bf k} + {\bf  K} / 2 - {\bf  Q})
u_j(-{\bf k} + {\bf  K}  / 2) \rangle
\nonumber\\
&\simeq& \biggl[i({\bf k } \cdot \bar{\bf B}) + \frac{1}{2}
(\bar{\bf B} \cdot \bec{\nabla})\biggr] f_{ij}({\bf k},{\bf R}) -
\frac{1}{2} k_{p} {\partial f_{ij}({\bf k}) \over
\partial k_s} \bar B_{p,s} \;
\nonumber\\
\label{B16}
\end{eqnarray}
(see \cite{RK2001}). We took into account that in Eq. (\ref{B8})
the terms with symmetric tensors with respect to the indexes "i"
and "j" do not contribute to the electromotive force because
${\cal E}_{m} = \varepsilon_{mji} \, g_{ij} $. In Eqs.
(\ref{B6})-(\ref{B8}) we neglected the second and higher
derivatives over $ {\bf R} .$ To derive Eqs. (\ref{B6})-(\ref{B8})
we also used the following identity
\begin{eqnarray}
&& i k_i \int f_{ij}({\bf k} - {1 \over 2}{\bf  Q}, {\bf K} - {\bf
Q}) \bar U_{p}({\bf  Q}) \exp(i {\bf K} {\bf \cdot} {\bf R}) \,d
{\bf  K} \,d {\bf  Q}
\nonumber\\
& & = -{1 \over 2} \bar U_{p} \nabla _i f_{ij} + {1 \over 2}
f_{ij} \nabla _i \bar U_{p} -  {i \over 4} (\nabla _s \bar U_{p})
\biggl(\nabla _i {\partial f_{ij} \over \partial k_s} \biggr)
\nonumber\\
& & +  {i \over 4} \biggl( {\partial f_{ij} \over
\partial k_s} \biggr) (\nabla _s \nabla _i \bar U_{p}) \;
\label{D100}
\end{eqnarray}
(see \cite{RK03}). We split the tensor of magnetic fluctuations
into nonhelical, $h_{ij},$ and helical, $h_{ij}^{(H)},$ parts. The
helical part of the tensor of magnetic fluctuations depends on the
magnetic helicity and it is not determined by Eq.~(\ref{B7}). The
tensor $h_{ij}^{(H)}$ is determined by the dynamic equation which
follows from the magnetic helicity conservation arguments
\cite{KR82,ZRS83} (see also
\cite{GD94,KRR95,KR99,KMRS2000,KMRS02,BB02}).

First, we consider a nonrotating and shear free turbulence $({\bf
\Omega} = 0; \, \, \nabla_i \bar{\bf U} = 0),$ and we omit tensors
$I_{ijmn}^\sigma(\bar{\bf U})$, $\, E_{ijmn}^\sigma(\bar{\bf U})$
and $J_{ijmn}^\sigma(\bar{\bf U})$ in Eqs.~(\ref{B6})-(\ref{B8}).
First we solve Eqs.~(\ref{B6})-(\ref{B8}) neglecting the sources
$I^f_{ij}, I^h_{ij}, I^g_{ij}$ with the large-scale spatial
derivatives. Then we will take into account the terms with the
large-scale spatial derivatives by perturbations. We start with
Eqs.~(\ref{B6})-(\ref{B8}) written for nonhelical parts of the
tensors, and then consider Eqs.~(\ref{B6})-(\ref{B8}) for helical
parts of the tensors.

We subtract Eqs. (\ref{B6})-(\ref{B8}) written for background
turbulence (for $\bar{\bf B}=0)$ from those for $\bar{\bf B}
\not=0$, use the $\tau$ approximation [which is determined by Eqs.
(\ref{A1})-(\ref{A3})], neglect the terms with the large-scale
spatial derivatives, assume that $\eta k^2 \ll \tau^{-1}(k)$ and
$\nu k^2 \ll \tau^{-1}(k)$ for the inertial range of turbulent
fluid flow, and assume that the characteristic time of variation
of the mean magnetic field $\bar{\bf B}$ is substantially larger
than the correlation time $\tau(k)$ for all turbulence scales. We
split all correlation functions into symmetric and antisymmetric
parts with respect to the wave number ${\bf k}$, {\em e.g.,} $
f_{ij} = f_{ij}^{(s)} + f_{ij}^{(a)} ,$ where $ f_{ij}^{(s)} =
[f_{ij}({\bf k}) + f_{ij}(-{\bf k})] / 2 $ is the symmetric part
and $ f_{ij}^{(a)} = [f_{ij}({\bf k}) - f_{ij}(-{\bf k})] / 2 $ is
the antisymmetric part, and similarly for other tensors. Thus, we
obtain
\begin{eqnarray}
\hat f_{ij}^{(s)}({\bf k}) &\approx& {1 \over 1 + 2 \psi} [(1 +
\psi) f_{ij}^{(0,s)}({\bf k}) + \psi h_{ij}^{(0,s)}({\bf k})]  \;,
\nonumber\\
\label{B22} \\
\hat h_{ij}^{(s)}({\bf k}) &\approx& {1 \over 1 + 2 \psi} [\psi
f_{ij}^{(0,s)}({\bf k}) + (1 + \psi) h_{ij}^{(0,s)}({\bf k})] \;,
\nonumber\\
\label{B24}\\
\hat g_{ij}^{(a)}({\bf k}) &\approx& {i \tau ({\bf k} {\bf \cdot}
\bar{\bf B}) \over 1 + 2 \psi} [f_{ij}^{(0,s)}({\bf k}) -
h_{ij}^{(0,s)}({\bf k})]  \; \label{B26}
\end{eqnarray}
(see \cite{RK04}), where $\hat f_{ij}, \hat h_{ij}$ and $\hat
g_{ij}$ are solutions without the sources $I^f_{ij}, I^h_{ij}$ and
$I^g_{ij}$, $\, \psi({\bf k}) = 2 (\tau \, {\bf k} {\bf \cdot}
\bar{\bf B})^2 .$ The correlation functions $\hat
f_{ij}^{(a)}({\bf k})$, $\hat h_{ij}^{(a)}({\bf k})$ and $\hat
g_{ij}^{(s)}({\bf k})$ vanish if we neglect the large-scale
spatial derivatives, i.e., they are proportional to the
first-order spatial derivatives.

Now we take into account the large-scale spatial derivatives in
Eqs. (\ref{B6})-(\ref{B8}) by perturbations. Their effect
determines the following steady-state equations for the second
moments $\tilde f_{ij}$, $\tilde h_{ij}$ and $\tilde g_{ij}$:
\begin{eqnarray}
\tilde f_{ij}^{(a)}({\bf k}) &=& f_{ij}^{(0,a)}({\bf k}) + i \tau
({\bf k} {\bf \cdot} \bar{\bf B}) \tilde \Phi_{ij}^{(M,s)}({\bf
k}) + \tau I^f_{ij} \;,
\nonumber \\
\label{B28}\\
\tilde h_{ij}^{(a)}({\bf k}) &=& h_{ij}^{(0,a)}({\bf k}) - i \tau
({\bf k} {\bf \cdot} \bar{\bf B}) \tilde \Phi_{ij}^{(M,s)}({\bf
k}) + \tau I^h_{ij} \;,
\nonumber \\
\label{B29} \\
\tilde g_{ij}^{(s)}({\bf k }) &=& i \tau ({\bf k} {\bf \cdot}
\bar{\bf B}) (\tilde f_{ij}^{(a)}({\bf k}) - \tilde
h_{ij}^{(a)}({\bf k})) + \tau I^g_{ij} \;,
\nonumber \\
\label{B30}
\end{eqnarray}
where $ \tilde \Phi_{ij}^{(M,s)} = [\tilde \Phi_{ij}^{(M)}({\bf
k}) + \tilde \Phi_{ij}^{(M)}(-{\bf k})] / 2 .$ Here $\tilde
f_{ij}$, $\tilde h_{ij}$ and $\tilde g_{ij}$ denote the
contributions to the second moments caused by the large-scale
spatial derivatives.  The correlation functions of the background
turbulence $f_{ij}^{(0,a)}({\bf k})$ and $h_{ij}^{(0,a)}({\bf k})$
are determined by the inhomogeneity of turbulence [see
Eqs.~(\ref{K1}) and~(\ref{K2})]. The solution of Eqs.
(\ref{B28})-(\ref{B30}) yield
\begin{eqnarray}
&& \tilde \Phi_{mn}^{(M,s)}({\bf k}) = {2 i \tau ({\bf k} {\bf
\cdot} \bar{\bf B}) \over 1 + 2 \psi} (f_{mn}^{(0,a)} -
h_{mn}^{((0,a))}) + \biggl[(1 + \epsilon)
\nonumber\\
& & \times (1 + 2 \psi) (\delta_{nj} \delta_{mk} - \delta_{mj}
\delta_{nk} + k_{nk} \delta_{mj} - k_{mk} \delta_{nj})
\nonumber\\
& & - 2 (\epsilon + 2 \psi) (k_{nj} \delta_{mk} - k_{mj}
\delta_{nk}) \biggr] {\tau \, \bar B_{j,k} \over (1 + 2 \psi)^2}
\; . \label{BB40}
\end{eqnarray}
The correlation functions $\tilde f_{ij}^{(s)}({\bf k})$, $\tilde
h_{ij}^{(s)}({\bf k})$ and $\tilde g_{ij}^{(a)}({\bf k})$ are of
the order of $\sim O(\nabla^2)$, i.e., they are proportional to
the second-order spatial derivatives. Thus $\hat f_{ij} + \tilde
f_{ij} $ is the nonhelical part of the correlation function of the
velocity field for a nonrotating turbulence, and similarly for
other second moments.

Next, we solve Eqs.~(\ref{B6})-(\ref{B8}) for helical parts of the
tensors for a nonrotating turbulence using the same approach which
we used before (see also \cite{RK04}). The steady-state solution
of Eqs.~(\ref{B6}) and (\ref{B8}) for the helical parts of the
tensor reads:
\begin{eqnarray}
\Phi_{ij}^{(M,H)}({\bf k}) &=& {2 i \tau ({\bf k} {\bf \cdot}
\bar{\bf B}) \over 1 + \psi} (f_{ij}^{(0,H)} - h_{ij}^{(H)}) \; .
\label{B34}
\end{eqnarray}
where $\Phi_{ij}^{(M,H)}({\bf k}) = g_{ij}^{(H)}({\bf k}) -
g_{ji}^{(H)}(-{\bf k}) $ and $f_{ij}^{(0,H)}({\bf k})$ is the
helical part of the tensor for velocity field of the background
turbulence. The tensor $h_{ij}^{(H)}$ is determined by the dynamic
equation \cite{KR82,KR99}. Since $f_{ij}^{(0,H)}$ and
$h_{ij}^{(H)}$ are of the order of $O(\nabla)$ we do not need to
take into account the source terms with the large-scale spatial
derivatives.

Now we determine the nonlinear electromotive force $ {\cal
E}_{i}({\bf r}=0) = (1/2) \varepsilon_{inm} \int
[\Phi_{mn}^{(M,H)}({\bf k}) + \tilde \Phi_{mn}^{(M,s)}({\bf k})]
\,d {\bf k} $ in a nonrotating and shear free turbulence:
\begin{eqnarray}
{\cal E}_{i} &=& \varepsilon_{inm} \int \biggl[ {i \tau ({\bf k}
{\bf \cdot} \bar{\bf B}) \over 1 + \psi} (f_{mn}^{(0,H)} -
h_{mn}^{(H)}) + {\tau \over 1 + 2 \psi} \{I^g_{mn}
\nonumber\\
& & + i({\bf k} {\bf \cdot} \bar{\bf B}) [f_{mn}^{(0,a)} -
h_{mn}^{(0,a)} + \tau (I^f_{mn} - I^h_{mn})]\} \biggr] \,d {\bf k}
\; .
\nonumber\\
\label{B35}
\end{eqnarray}
To integrate in $ {\bf k} $-space in the nonlinear electromotive
force we specify a model for the background turbulence [see Eqs.
(\ref{K1})-(\ref{K2})]. After the integration in $ {\bf k} $-space
we obtain the nonlinear electromotive force:
\begin{eqnarray}
{\cal E}_{i} = a_{ij} \bar B_{j} + b_{ijk} \bar B_{j,k} \;,
\label{A23}
\end{eqnarray}
where $ \bar B_{i,j} = \partial \bar B_{i} / \partial R_{j} ,$
$\varepsilon_{ijk}$ is the Levi-Civita tensor, and the tensors
$a_{ij}$ and $b_{ijk}$ are given by
\begin{eqnarray}
a_{ij} &=& {1 \over 6} \tau_0 \biggl[A_{1}^{(1)}(\sqrt{2} \beta)
\, \varepsilon_{ijn} - A_{2}^{(1)}(\sqrt{2} \beta)
\varepsilon_{inm} \, \beta_{mj} \biggr]
\nonumber\\
&& \times \nabla_n [\langle {\bf u}^2 \rangle^{(0)} - \langle {\bf
b}^2 \rangle^{(0)}] + [\chi^{v} \, \phi^{v}(\beta)
\nonumber\\
&& + \chi^{c}(\bar{\bf B}) \, \phi^{m}(\beta)] \, \delta_{ij} \;,
\label{L1}\\
b_{ijk} &=& \eta_{_{T}} \biggl[\phi_1(\bar{B}) \,
\varepsilon_{ijk} + \phi_2(\bar{B}) \, \varepsilon_{ijn} \,
\beta_{nk}
\nonumber\\
&& + \phi_3(\bar{B}) \, \varepsilon_{ink} \, \beta_{nj} \biggr]
\;, \label{L2}
\end{eqnarray}
where $\beta_{ij} = \bar B_{i} \bar B_{j} / \bar B^2$, the
quenching functions $\phi^v(\beta)$, $\, \phi^m(\beta)$ and
$\phi_k(\bar{B})$ are determined by Eqs.~(\ref{X22}), (\ref{X23})
and (\ref{LLL3}), respectively, $\beta = 4 \bar B / (u_{0} \sqrt{2
\mu \rho}) ,$ $\, \epsilon = \langle {\bf b}^2 \rangle^{(0)} /
\langle {\bf u}^2 \rangle^{(0)}$, and all calculations are made
for $q=5/3$. The parameter $\chi^{v} = - \tau_0 \, \mu^{v} / 3$ is
related to the hydrodynamic helicity $\mu^{v}$ of the background
turbulence, and the function $\chi^{c}(\bar {\bf B}) = (\tau / 3
\mu \rho) \langle {\bf b} \cdot (\bec{\nabla} {\bf \times} {\bf
b}) \rangle$ is related to the current helicity. These parameters
are written in the dimensional form. To integrate over the angles
in $ {\bf k} $-space we used the following identity:
\begin{eqnarray}
\bar K_{ij} = \int {k_{ij} \sin \theta \over 1 + a \cos^{2}
\theta} \,d\theta \,d\varphi = \bar A_{1} \delta_{ij} + \bar A_{2}
\beta_{ij} \;,
\label{C22}
\end{eqnarray}
where $ a = \beta^2 / \bar \tau(k) ,$ and
\begin{eqnarray}
\bar A_{1} &=& {2 \pi \over a} \biggl[(a + 1) {\arctan (\sqrt{a})
\over \sqrt{a}} - 1 \biggr] \;,
\label{X24}\\
\bar A_{2} &=& - {2 \pi \over a} \biggl[(a + 3) {\arctan
(\sqrt{a}) \over \sqrt{a}} - 3 \biggr] \;
\label{X26}
\end{eqnarray}
(for details, see \cite{RK2001,RK04}). The functions
$A_{n}^{(1)}(\beta)$ are given by
\begin{eqnarray}
A_{n}^{(1)}(\beta) = {3 \beta^{4} \over \pi} \int_{\beta}^{\infty}
{\bar A_{n}(X^{2}) \over X^{5}} \,d X  \;,
\label{X27}
\end{eqnarray}
where $ X^{2} = \beta^{2} (k / k_{0})^{2/3} = a$, and we took into
account that the inertial range of the turbulence exists in the
scales: $ l_{d} \leq r \leq l_{0} .$ Here the maximum scale of the
turbulence $ l_{0} \ll L ,$ and $ l_{d} = l_{0} / {\rm Re}^{3/4} $
is the viscous scale of turbulence, ${\rm Re} = l_{0} u_{0} / \nu$
is the Reynolds number, $\nu$ is the kinematic viscosity and $ L $
is the characteristic scale of variations of the nonuniform mean
magnetic field. For very large Reynolds numbers $ k_{d} =
l_{d}^{-1} $ is very large and the turbulent hydrodynamic and
magnetic energies are very small in the viscous dissipative range
of the turbulence $ 0 \leq r   \leq l_{d} .$ Thus we integrated in
$ \bar A_{n} $ over $ k $ from $ k_{0} = l_{0}^{-1} $ to $ \infty
.$ We also used the following identity $\int_0^1 \bar A_{n}(a(\bar
\tau)) \bar \tau \, d\bar \tau = (2 \pi / 3) A_{n}^{(1)}(\beta)$.
The explicit form of the functions $\bar A_{k}(\beta^2)$ and
$A_{k}^{(1)}(\beta)$, and their asymptotic formulas are given in
\cite{RK04}.

We use an identity $ \bar B_{j,i} = (\partial \hat B)_{ij} +
\varepsilon_{ijn} (\bec{\bf \nabla} {\bf \times} \bar{\bf B})_{n}
/ 2 $ which allows us to rewrite Eq. (\ref{A23}) for the
electromotive force in the form of Eq. (\ref{A14}), where
\begin{eqnarray}
\alpha_{ij}(\bar{\bf B}) &=& {1 \over 2} (a_{ij} + a_{ji}) \;,
\quad V^{\rm eff}_k(\bar{\bf B}) = {1 \over 2} \varepsilon_{kji}
a_{ij} \;,
\label{A25} \\
\eta_{ij}(\bar{\bf B}) &=& {1 \over 4} (\varepsilon_{ikp} b_{jkp}
+ \varepsilon_{jkp} b_{ikp}) \;,
\label{A26}  \\
\delta_{i} &=& {1 \over 4} (b_{jji} - b_{jij}) \;, \quad
\kappa_{ijk}(\bar{\bf B}) = - {1 \over 2} (b_{ijk} + b_{ikj}) .
\nonumber\\
\label{A27}
\end{eqnarray}
(see \cite{R80}). Using Eqs.~(\ref{A25})-(\ref{A27})
and~(\ref{L1})-(\ref{L2}) we derive equations for the coefficients
defining nonlinear electromotive force for a nonrotating
turbulence. In particular,
\begin{eqnarray}
\alpha_{ij}(\bar{\bf B}) &=& [\chi^{v} \, \phi^{v}(\beta) +
\chi^{c}(\bar{\bf B}) \, \phi^{m}(\beta)]  \, \delta_{ij} \;,
\label{BL1}\\
{\bf V}^{\rm eff} &=& \eta_{_{T}} \, \biggl[{\bf V}_{d}(\bar{B}) -
{\phi_3(\bar{B}) \over 2}  \, {\bf \Lambda}^{(B)}
\nonumber\\
&& + {\phi_2(\bar{B}) \over \bar{\bf B}^2} \, (\bar{\bf B} \cdot
\bec{\nabla}) \bar{\bf B} \biggr] \;,
\label{B40}\\
\eta_{ij} &=& \eta_{_{T}} \phi_1(\bar{B}) \, \delta_{ij} \;,
\label{B41}
\end{eqnarray}
where
\begin{eqnarray}
{\bf V}_{d}(\bar{B}) &=& - {1 \over 2} \phi_1(\bar{B}) ({\bf
\Lambda}^{(u)} - \epsilon {\bf \Lambda}^{(b)}) \;, \label{LL5}
\end{eqnarray}
and ${\bf \Lambda}^{(B)} = (\bec{\nabla} \bar{\bf B}^2) / \bar{\bf
B}^2 $. Note that Eqs.~(\ref{B28})-(\ref{L2}) and
(\ref{BL1})-(\ref{B41}) for a homogeneous and nonhelical
background turbulence coincide with those derived in \cite{RK04}.

Now we study the effect of a mean uniform rotation of the fluid on
the nonlinear electromotive force in a shear free turbulence. We
consider a slow rotation rate $(\tau \Omega \ll 1) ,$ i.e., we
neglect terms $ \sim O(\Omega^2) .$ We also neglect terms $ \sim
O(\nabla^2) .$ However, we take into account terms $ \sim
O(\Omega_i \nabla_j) ,$ that is possible by the following symmetry
reasons. The tensor $\Omega_i \nabla_j$ is a pseudo tensor, while
$\Omega_i \Omega_j$ and $\nabla_i \nabla_j$ are true tensors. This
implies that a pseudo tensor quantity includes terms $ \propto
\Omega_i \nabla_j ,$ but does not include terms $ \propto \Omega_i
\Omega_j$ and $ \propto \nabla_i \nabla_j$. On the other hand, a
true tensor quantity does not include terms $ \propto \Omega_i
\nabla_j $, but it may include the terms $ \propto \Omega_i
\Omega_j$ and $ \propto \nabla_i \nabla_j$. The steady-state
solution of Eqs.~(\ref{B6}) and (\ref{B8}) for the nonhelical
parts of the tensors for a rotating turbulence reads:
\begin{eqnarray}
N_{ijpq}^f({\bf \Omega}) f_{pq} &=& \tau \{ i({\bf k} {\bf \cdot}
\bar{\bf B}) \Phi_{ij}^{(M)} + I^f_{ij} \} \;,
\label{S1} \\
N^g_{nj}({\bf \Omega}) g_{in} &=& \tau \{ i({\bf k} {\bf \cdot}
\bar{\bf B}) [f_{ij}({\bf k}) - h_{ij}({\bf k})] + I^g_{ij} \} \;,
\nonumber \\
\label{S2}
\end{eqnarray}
where $ N_{ijpq}^f({\bf \Omega}) = \delta_{ip} \delta_{jq} - \tau
M_{ijpq}^\Omega $ and $ N_{ij}^g({\bf \Omega}) = \delta_{ij} -
\tau D_{ij}^\Omega .$ Here we use the following notations: the
total correlation function is $f_{ij} = \bar f_{ij} +
f_{ij}^{\Omega}$, where $ \bar f_{ij} = \hat f_{ij} + \tilde
f_{ij} $ is the correlation functions for a nonrotating
turbulence, and $f_{ij}^{\Omega}$ determines the contribution to
the correlation function of the velocity field caused by a uniform
rotation. The similar notations are for other correlation
functions. Now we solve Eqs.~(\ref{B7}), (\ref{S1}) and (\ref{S2})
by iteration which yields
\begin{eqnarray}
f_{ij}^{\Omega}({\bf k}) &=& \tau \{M_{ijpq}^\Omega \bar f_{pq} +
i({\bf k} {\bf \cdot} \bar{\bf B}) \Phi_{ij}^{(M,\Omega)}
\nonumber\\
& & + I^{f,\Omega}_{ij}(g_{ij}^{\Omega}) \} \;,
\label{S3} \\
h_{ij}^{\Omega}({\bf k}) &=& - \tau \{i({\bf k} {\bf \cdot}
\bar{\bf B}) \Phi_{ij}^{(M,\Omega)} -
I^{h,\Omega}_{ij}(g_{ij}^{\Omega}) \} \;,
\label{S4} \\
g_{ij}^{\Omega}({\bf k}) &=& \tau \{D_{jn}^\Omega \bar g_{in} + i
({\bf k} {\bf \cdot} \bar{\bf B}) [f_{ij}^{\Omega} -
h_{ij}^{\Omega}]
\nonumber\\
& & + I^{g,\Omega}_{ij}(f_{ij}^{\Omega}, h_{ij}^{\Omega}) \} \;,
\label{S5}
\end{eqnarray}
where $\Phi_{ij}^{(M,\Omega)}({\bf k}) = g_{ij}^{\Omega}({\bf k})
- g_{ji}^{\Omega}(-{\bf k}) $, the source terms
$I^{f,\Omega}_{ij}(g_{ij}^{\Omega})$, $\,
I^{h,\Omega}_{ij}(g_{ij}^{\Omega})$ and
$I^{g,\Omega}_{ij}(f_{ij}^{\Omega}, h_{ij}^{\Omega})$ are
determined by Eqs.~(\ref{MM1})-(\ref{MM3}), where $f_{ij}$,
$h_{ij}$, $g_{ij}$ are replaced by $f_{ij}^{\Omega}$,
$h_{ij}^{\Omega}$, $g_{ij}^{\Omega}$, respectively. The solution
of Eqs. (\ref{S3})-(\ref{S5}) yield equation for the symmetric
part $\Phi_{ij}^{(M,\Omega,s)}$ of the tensor:
\begin{eqnarray}
\Phi_{ij}^{(M,\Omega,s)}({\bf k}) &=& {\tau \over 1 + 2 \psi} \{
D_{jn}^\Omega \bar g_{in} - D_{in}^\Omega \bar g_{jn} + i \tau
({\bf k} {\bf \cdot} \bar{\bf B}) [I^{f,\Omega}_{ij}
\nonumber\\
& & - I^{f,\Omega}_{ji} + I^{h,\Omega}_{ji} - I^{h,\Omega}_{ij} +
2 M_{ijpq}^\Omega \bar f_{pq}]
\nonumber\\
& & + I^{g,\Omega}_{ij} - I^{g,\Omega}_{ji} \} \; .
\label{S6}
\end{eqnarray}
Thus, the effect of a uniform rotation on the nonlinear
electromotive force, $ {\cal E}^\Omega_{i}({\bf r}=0) \equiv (1/2)
\varepsilon_{inm} \int \Phi_{mn}^{(M,\Omega,s)} \,d {\bf k} ,$ is
determined by
\begin{eqnarray}
{\cal E}^\Omega_{i} &=& \varepsilon_{inm} \int {\tau \over 1 + 2
\psi} \{D_{np}^\Omega \bar g_{mp} + i \tau ({\bf k} {\bf \cdot}
\bar{\bf B}) [M_{mnpq}^\Omega \bar f_{pq}
\nonumber \\
& & + I^{f,\Omega}_{mn} - I^{h,\Omega}_{mn}] + I^{g,\Omega}_{mn}\}
\, d{\bf k} \; .
\label{S7}
\end{eqnarray}
Now we use the following identities:
\begin{eqnarray*}
\varepsilon_{inm} \hat D_{np}^\Omega \tilde g_{mp} &=& 2 \Omega_m
(k_{im} \tilde g_{pp} - k_{nm} \tilde g_{ni}) \;,
\\
\varepsilon_{inm} \tilde D_{np}^\Omega \hat g_{mp} &=& 2 i (T_{i}
\hat g_{pp} - T_{n} \hat g_{ni}) \;,
\\
\varepsilon_{inm} \hat M_{mnpq}^\Omega \tilde f_{pq} &=& 2
\Omega_m k_{nm} (\tilde f_{in} - \tilde f_{ni}) \;,
\\
i \varepsilon_{inm} \tilde M_{mnpq}^\Omega \hat f_{pq} &=& 4
(T_{i} \hat f_{pp} - T_{n} \hat f_{in}) \;,
\end{eqnarray*}
where $T_{i} = \Omega_m T_{mip} \nabla_p .$ We also take into
account that
\begin{eqnarray*}
k_n \tilde f_{ni} &=& (i/2) \nabla_n \hat f_{ni} \;, \quad k_n
\tilde f_{in} = - (i/2) \nabla_n \hat f_{in}  \;,
\\
k_n \tilde g_{ni} &=& (i/2) \nabla_n \hat g_{ni} \; .
\end{eqnarray*}
These equations follow from the condition $\bec{\nabla} {\bf
\cdot} {\bf u} = 0 .$ Thus we obtain that the effect of  a uniform
rotation on the nonlinear electromotive force is determined by
${\cal E}_{i}^\Omega \equiv a_{ij}^\Omega \bar B_{j} +
b_{ijk}^\Omega \bar B_{j,k} $, where
\begin{eqnarray}
a_{ij}^\Omega &=& \int {2 \tau^2 \Omega_m \over 1 + 2 \psi} \,
\nabla_k \, \biggl\{(k_{ijmk} - k_{ij} \, \delta_{mk}) \biggl[2
\langle {\bf u}^2 \rangle^{(0)}
\nonumber \\
& & + \, {1 - 2 \psi \over 1 + 2 \psi} \, [\langle {\bf u}^2
\rangle^{(0)} - \langle {\bf b}^2 \rangle^{(0)}] \biggr]
\nonumber \\
& & + \, k_{im} \, \delta_{jk} \, [\langle {\bf u}^2 \rangle^{(0)}
+ \langle {\bf b}^2 \rangle^{(0)}] \biggr\} \, d{\bf k} \;,
\label{S8}\\
b_{ijk}^\Omega &=& \int {2 \tau^2 \Omega_m \over (1 + 2 \psi)^2}
\biggl\{ \biggl[2(\psi - 1) + {4 \over 1 + 2 \psi} \biggr] \,
k_{ijmk}
\nonumber \\
& & - 2\psi \, k_{jm} \, \delta_{ik} +  \biggl[3 - {4 \over 1 + 2
\psi} \biggr] \, k_{ij} \, \delta_{mk} \, [\langle {\bf u}^2
\rangle^{(0)}
\nonumber \\
& & - \langle {\bf b}^2 \rangle^{(0)}] - (1 + 2 \psi) \, k \,
k_{ijm} \, {\partial \over \partial k_k} \, [\langle {\bf u}^2
\rangle^{(0)}
\nonumber \\
& & + \langle {\bf b}^2 \rangle^{(0)}] \biggr\} \, d{\bf k} \;,
\label{S9}
\end{eqnarray}
and we used the identities:
\begin{eqnarray*}
({\bf k} {\bf \cdot} \bar{\bf B}) \nabla_n \psi &=& 2 \psi \, k_j
\, \bar B_{j,n} \;,
\\
{\partial \psi \over \partial k_i} &=& 4 \tau^2 \, ({\bf k} {\bf
\cdot} \bar{\bf B}) \, \bar B_{i} - 2(q-1) \psi \, {k_i \over k^2}
\; .
\end{eqnarray*}
Now we use the following identities:
\begin{eqnarray*}
&& \bar B_{j} \bar K_{ijmn} \Omega_m \Lambda_n = \{ (\bar C_{1} +
\bar C_{2}) \, [\Omega_i \Lambda_j + \Omega_j \Lambda_i
\\
&& \quad \quad + \, \delta_{ij} \, ({\bf \Omega} {\bf \cdot} {\bf
\Lambda})] + (\bar C_{2} + 3 \bar C_{3}) \, \delta_{ij} \, ({\bf
\Omega} {\bf \cdot} \bec{\hat \beta}) ({\bf \Lambda} {\bf \cdot}
\bec{\hat \beta}) \} \bar B_{j} \;,
\\
&& \bar K_{ij} \Omega_j (\bar{\bf B} {\bf \cdot} {\bf \Lambda}) =
[\bar A_{1} \, \Omega_i \Lambda_j + \bar A_{2} \, \delta_{ij} \,
({\bf \Omega} {\bf \cdot} \bec{\hat \beta}) ({\bf \Lambda} {\bf
\cdot} \bec{\hat \beta}) ] \bar B_{j} \;,
\\
&& \bar B_{j} \bar K_{ij} = (\bar A_{1} + \bar A_{2}) \bar B_{i}
\;,
\\
&& \bar B_{j,i} \, \bar K_{jm} \Omega_m = \bar A_{1} \, \nabla_i
({\bf \Omega} {\bf \cdot} \bar{\bf B}) + {1 \over 2} \, \bar A_{2}
\, \Lambda^{(B)}_i ({\bf \Omega} {\bf \cdot} \bar{\bf B}) \;,
\\
&& \bar B_{j,m} \, \bar K_{ij} \Omega_m = [\bar A_{1} \, ({\bf
\Omega} {\bf \cdot} \bec{\nabla}) + {1 \over 2} \, \bar A_{2} \,
({\bf \Omega} {\bf \cdot} {\bf \Lambda}^{(B)})] \, \bar B_{i} \;,
\end{eqnarray*}
\begin{widetext}
\begin{eqnarray*}
&& \bar B_{j,k} \, \bar K_{ijmk} \Omega_m = \bar C_{1} \,
(\Omega_j \, \delta_{ik} + \Omega_k \, \delta_{ij}) \, \bar
B_{j,k} + {1 \over 2} \, \biggl\{ \bar C_{3} \, \, [2
\Lambda^{(B)}_i ({\bf \Omega} {\bf \cdot} \bar{\bf B}) + \Omega_i
({\bf \Lambda}^{(B)} {\bf \cdot} \bar{\bf B})] + \bar C_{3} \,
\bar B_{i} \, \biggl[({\bf \Omega} {\bf \cdot} {\bf
\Lambda}^{(B)})
\\
&& \quad \quad +  \, {2 \over \bar B} (\bec{\hat \beta} {\bf
\cdot} \bec{\nabla}) ({\bf \Omega} {\bf \cdot} \bar{\bf B})
\biggr] + \bar C_{2} \, ({\bf \Omega} {\bf \cdot} \bec{\hat
\beta}) ({\bf \Lambda}^{(B)} {\bf \cdot} \bec{\hat \beta}) \, \bar
B_{i} \biggr \} -  \, \bar C_{3} \, {1 \over \bar B} \, ({\bf
\Omega} {\bf \cdot} \bec{\hat \beta}) \, (\bar{\bf B} {\bf \times}
(\bec{\nabla} {\bf \times} \bar{\bf B})) \;,
\\
&& \bar B_{j} \, (\bar{\bf B} {\bf \cdot} \bec{\nabla}) \, \bar
B_{n} \, \bar K_{ijmn} \Omega_m = {1 \over 2} \, \biggl[ (\bar
C_{1} + \bar C_{3}) \, \{ [\Lambda^{(B)}_i ({\bf \Omega} {\bf
\cdot} \bar{\bf B}) + \, \Omega_i ({\bf \Lambda}^{(B)} {\bf \cdot}
\bar{\bf B}) + \bar B_{i} ({\bf \Omega} {\bf \cdot} {\bf
\Lambda}^{(B)})] \bar B^2 - 2 [\Omega_j \bar B_{i}
\\
&& \quad \quad + \,  \delta_{ij} ({\bf \Omega} {\bf \cdot}
\bar{\bf B})] (\bar{\bf B} {\bf \times} (\bec{\nabla} {\bf \times}
\bar{\bf B}))_j \} + (\bar C_{2} + 3 \bar C_{3}) \, ({\bf \Omega}
{\bf \cdot} \bar{\bf B}) ({\bf \Lambda}^{(B)} {\bf \cdot} \bar{\bf
B}) \biggr] \;,
\end{eqnarray*}
where
\begin{eqnarray}
\bar K_{ijmn} &=& \int {k_{ijmn} \sin \theta \over 1 + a \cos^{2}
\theta} \,d\theta \,d\varphi = \bar C_{1} (\delta_{ij} \delta_{mn}
+ \delta_{im} \delta_{jn} + \delta_{in} \delta_{jm}) + \bar C_{2}
\beta_{ijmn} + \bar C_{3} (\delta_{ij} \beta_{mn} + \delta_{im}
\beta_{jn}
\nonumber \\
&&+ \delta_{in} \beta_{jm} + \delta_{jm} \beta_{in} + \delta_{jn}
\beta_{im} + \delta_{mn} \beta_{ij}) \;,
\label{C24}
\end{eqnarray}
\bigskip
\end{widetext}
\noindent
and
\begin{eqnarray*}
\bar C_{1} &=& {\pi \over 2a^{2}} \biggl[(a + 1)^{2} {\arctan
(\sqrt{a}) \over \sqrt{a}} - {5 a \over 3} - 1 \biggr]  \;,
\\
\bar C_{2} &=& \bar A_{2} - 7 \bar A_{1} + 35 \bar C_{1} \;, \quad
\bar C_{3} = \bar A_{1} - 5 \bar C_{1} \; .
\end{eqnarray*}
Integration in ${\bf k}$-space yields
\begin{widetext}
\begin{eqnarray}
a_{ij}^\Omega &=& {2 \over 3} l_0^2 \biggl\{ E_3 \, \Omega_j
\Lambda^{(B)}_i + E_4 \, \Omega_i \Lambda^{(B)}_j + {1 \over \bar
B^2} \biggl[\delta_{ij} \, \biggl(E_5 \, ({\bf \Omega} {\bf \cdot}
{\bf \Lambda}^{(B)}) \bar B^2 + [E_6 \, (\bar{\bf B} {\bf \cdot}
{\bf \Lambda}^{(B)}) + \, E_7 \, (\bar{\bf B} {\bf \cdot}
\bec{\nabla})] ({\bf \Omega} {\bf \cdot} \bar{\bf B})
\nonumber \\
&& + E_8 \, {\bf \Omega} {\bf \cdot} (\bar{\bf B} {\bf \times}
(\bec{\nabla} {\bf \times} \bar{\bf B})) \biggr) - E_9 \,
\varepsilon_{ijm} (\bec{\nabla} {\bf \times} \bar{\bf B})_m ({\bf
\Omega} {\bf \cdot} \bar{\bf B}) \biggr] + E_{10} \, \Omega_j
\Lambda^{(v)}_i + E_{11} \, \Omega_i \Lambda^{(v)}_j + \delta_{ij}
\, \biggl(E_{12} \, ({\bf \Omega} {\bf \cdot} {\bf \Lambda}^{(v)})
\nonumber \\
&& + E_{13} \, {1 \over \bar B^2} ({\bf \Omega} {\bf \cdot}
\bar{\bf B}) (\bar{\bf B} {\bf \cdot} {\bf \Lambda}^{(v)}) \biggr)
+ \epsilon \biggl[E_{14} \, \Omega_j \Lambda^{(b)}_i + E_{15} \,
\Omega_i \Lambda^{(b)}_j + \delta_{ij} \, \biggl(E_{16} \, ({\bf
\Omega} {\bf \cdot} {\bf \Lambda}^{(b)}) + E_{17} \, {1 \over \bar
B^2} ({\bf \Omega} {\bf \cdot} \bar{\bf B}) (\bar{\bf B} {\bf
\cdot} {\bf \Lambda}^{(b)}) \biggr) \biggl] \biggr\} \;,
\nonumber \\
\label{S10}\\
b_{ijk}^\Omega &=& {2 \over 3} l_0^2 (E_1 \, \Omega_j \delta_{ik}
+ E_2 \, \Omega_k \delta_{ij}) \;,
\label{S11}
\end{eqnarray}
\end{widetext}
\noindent
where
\begin{eqnarray}
\Psi_1\{X\}_{y} &=& 3 X^{(1)}(y) - {3 \over 2 \pi} \bar X(y^2) \;,
\nonumber \\
\Psi_2\{X\}_{y} &=& 4 X^{(2)}(y) - {3 \over 2 \pi} \bar X(y^2) \;,
\nonumber \\
\Psi_3\{X\}_{y} &=& 6 X^{(2)}(y) - {3 \over \pi} \bar X(y^2) + {3
\over 4 \pi} y^2 \bar X'_{z=y^2} \;,
\nonumber \\
\Psi_4\{X\}_{y} &=& [2 - (1 + \epsilon) (2q -1)] \Psi_2\{X\}_{y}
\nonumber \\
&& + 4(1 - \epsilon) \Psi_3\{X\}_{y} - (1 + 3 \epsilon) X^{(2)}(y)
\;,
\nonumber \\
\Psi_5\{X\}_{y} &=& 2 \Psi_2\{X\}_{y} + X^{(2)}(y) \;,
\nonumber \\
\Psi_6\{X\}_{y} &=& - 2 \Psi_2\{X\}_{y} + X^{(2)}(y) \;,
\label{X1}
\end{eqnarray}
$\bar X' = d\bar X / dz$, and all calculations are made for
$q=5/3$,
\begin{eqnarray*}
E_1 &=& \biggl[\Psi_4\{C_1\}_{y} + (1 - \epsilon)
\Psi_2\{A_1\}_{y} + 2 \epsilon A_1^{(2)}(y) \biggr]_{y=\sqrt{2}
\beta},
\\
E_2 &=& \biggl[\Psi_4\{C_1\}_{y} - (1 - \epsilon) (\Psi_2 + 4
\Psi_3)\{A_1\}_{y}
\\
&& + (1 + \epsilon) A_1^{(2)}(y) \biggr]_{y=\sqrt{2} \beta} \;,
\\
E_3 &=& E_7 - {1 \over 2} E_8 + {1 \over 2} \, \biggl[(1 -
\epsilon) \Psi_2\{A_2\}_{y}
\\
&& + 2 \epsilon A_2^{(2)}(y) \biggr]_{y=\sqrt{2} \beta} \;,
\\
E_4 &=& {1 \over 2} (E_7 - E_8) \;,
\\
E_5 &=& E_4 + {1 \over 2} \, \biggl[(1 + \epsilon) \, A_2^{(2)}(y)
- (1 - \epsilon) (\Psi_2
\\
&& + 4 \Psi_3) \{A_2\}_{y} \biggr]_{y=\sqrt{2} \beta} \;,
\\
E_6 &=& {1 \over 2} \, \biggl[\Psi_4\{C_2\}_{y} - 2 (1 + \epsilon)
\, y^2 \, \Psi_1\{C_2 + 3C_3\}_{y} \biggr]_{y=\sqrt{2} \beta} \;,
\\
E_7 &=& \Psi_4\{C_3\}_{y=\sqrt{2} \beta} \;, \quad E_9 = E_7 - E_8
\;,
\\
E_8 &=& 4 (1 + \epsilon) \, \beta^2 \, \Psi_1\{C_1 +
C_3\}_{\sqrt{2} \beta} \;,
\\
E_{10} &=& \Psi_5\{C_1 + C_3\}_{\sqrt{2} \beta} \;, \quad E_{11} =
E_{10} + A_1^{(2)}(\sqrt{2} \beta) \;,
\\
E_{12} &=& E_{10} - \Psi_5\{A_1 + A_2\}_{\sqrt{2} \beta} \;,
\\
E_{13} &=& \biggl[\Psi_5\{C_2 + 3 C_3\}_{y} + A_2^{(2)}(y)
\biggr]_{y=\sqrt{2} \beta} \;,
\\
E_{14} &=& \Psi_6\{C_1 + C_3\}_{\sqrt{2} \beta} \;, \quad E_{15} =
E_{14} + A_1^{(2)}(\sqrt{2} \beta) \;,
\\
E_{16} &=& E_{14} - \Psi_6\{A_1 + A_2\}_{\sqrt{2} \beta} \;,
\\
E_{17} &=& \biggl[\Psi_6\{C_2 + 3 C_3\}_{y} + A_2^{(2)}(y)
\biggr]_{y=\sqrt{2} \beta} \; .
\end{eqnarray*}
Note that $\Psi_1\{A_{1}\}_y = A_{1}^{(1)}(y)  + (1/2)
A_{2}^{(1)}(y) .$ The functions $A_{n}^{(2)}(\beta)$ are given by
\begin{eqnarray}
A_{n}^{(2)}(\beta) = {3 \beta^{6} \over \pi} \int_{\beta}^{\infty}
{\bar A_{n}(X^{2}) \over X^{7}} \,d X \;,
\label{X40}
\end{eqnarray}
and similarly for $ C_{n}^{(2)}(\beta) .$ We used the following
identity $\int_0^1 \bar A_{n}(a(\bar \tau)) \bar \tau^2 \, d\bar
\tau = (2 \pi / 3) A_{n}^{(2)}(\beta)$, and similarly for $
C_{n}^{(2)}(\beta) .$ The explicit form of the functions
$A_{n}^{(k)}(\beta)$ and $C_{n}^{(k)}(\beta)$ and their asymptotic
formulas are given in \cite{RK04}.

The asymptotic formulas for the tensors $a_{ij}^\Omega$ and
$b_{ijk}^\Omega$ for a weak mean magnetic field $\bar{B} \ll
\bar{B}_{\rm eq} / 4$ are given by
\begin{eqnarray}
a_{ij}^\Omega &=& {2 \, \tau_0^2 \over 45} \, \biggl[(\Omega_i
\nabla_{j} + \Omega_j \nabla_{i}) (11 \, \langle {\bf u}^2
\rangle^{(0)} + 3 \, \langle {\bf b}^2 \rangle^{(0)})
\nonumber \\
&& - 8 \, \delta_{ij} \, ({\bf \Omega} {\bf \cdot} \bec{\nabla})
(3 \, \langle {\bf u}^2 \rangle^{(0)} - \langle {\bf b}^2
\rangle^{(0)})\biggr] \;,
\label{S12}\\
b_{ijk}^\Omega &=& {4 \, l_0^2 \over 135} \, \biggl[(11 -
\epsilon) \, \Omega_j \delta_{ik} - 2(2 - 7 \epsilon) \, \Omega_k
\delta_{ij}\biggr] \;,
\nonumber \\
\label{S14}
\end{eqnarray}
and for $\bar{B} \gg \bar{B}_{\rm eq} / 4$  they are given by
\begin{eqnarray}
a_{ij}^\Omega &\approx& -  {11 l_0^2 \over 3 \beta} \, \delta_{ij}
\, ({\bf \Omega} {\bf \cdot} {\bf \Lambda}^{(B)}) \, (1 - 1.3
\epsilon) - {\tau_0^2  \over 3 \beta^2} \, \biggl[\delta_{ij} \,
({\bf \Omega} {\bf \cdot} \bec{\nabla})
\nonumber \\
&& - {6 \pi \beta \over 7 \sqrt{2}} \, \Omega_i \nabla_{j} \biggr]
\, (\langle {\bf u}^2 \rangle + \langle {\bf b}^2 \rangle) \;,
\label{S15}\\
b_{ijk}^\Omega &\approx& - 3 {l_0^2 \over \beta} \, \biggl[(1 -
\epsilon) \, \Omega_j \delta_{ik} + 5(1 - \epsilon) \, \Omega_k
\delta_{ij} \biggr] \; .
\nonumber \\
\label{S16}
\end{eqnarray}
Using Eqs.~(\ref{A25})-(\ref{A27}) and~(\ref{S10})-(\ref{S11}) we
derive formulas for the contributions to the coefficients defining
the nonlinear electromotive force due to a uniform rotation. In
particular, the isotropic contribution to the hydrodynamic part of
the $\alpha$ effect caused by a uniform rotation is given by
\begin{eqnarray}
\alpha_{ij}^\Omega = \alpha^\Omega \, \delta_{ij} \;,
\label{CC4}
\end{eqnarray}
where $\alpha^\Omega$ is given by Eq.~(\ref{CCC4}), and the
quenching functions $\phi^\Omega_1(\bar{B})$ and
$\phi^\Omega_2(\bar{B})$ which determine $\alpha^\Omega$, are
given by
\begin{eqnarray}
\phi^\Omega_1(\bar{B}) &=& \Psi_5\{A_1 + A_2 - C_1 -
C_3\}_{\sqrt{2} \beta} \;,
\label{SS20}\\
\phi^\Omega_2(\bar{B}) &=& \Psi_6\{A_1 + A_2 - C_1 -
C_3\}_{\sqrt{2} \beta} \; . \label{SS21}
\end{eqnarray}
The coefficients defining the nonlinear electromotive force due to
a uniform rotation for a weak mean magnetic field $\bar{B} \ll
\bar{B}_{\rm eq} / 4$ are given by:
\begin{eqnarray}
\alpha_{ij}^\Omega &=& {2 \, \tau_0^2 \over 45} \,
\biggl[(\Omega_i \nabla_{j} + \Omega_j \nabla_{i}) (11 \, \langle
{\bf u}^2 \rangle^{(0)} + 3 \, \langle {\bf b}^2 \rangle^{(0)})
\nonumber \\
&& - 8 \, \delta_{ij} \, ({\bf \Omega} {\bf \cdot} \bec{\nabla})
(3 \, \langle {\bf u}^2 \rangle^{(0)} - \langle {\bf b}^2
\rangle^{(0)})\biggr] \;,
\label{SS12}\\
\bec{\delta}^\Omega &=& - {2 \, l_0^2 \over 9} (1 - \epsilon) {\bf
\Omega} \;,
\label{S17} \\
\kappa_{ijk}^\Omega &=& - {14 \, l_0^2 \over 135} \biggl[1 + {13
\over 7} \epsilon \biggr] (\Omega_j \delta_{ik} + \Omega_k
\delta_{ij}) \;, \label{S18}
\end{eqnarray}
and for $\bar{B} \gg \bar{B}_{\rm eq} / 4$ they are given by
\begin{eqnarray}
\alpha_{ij}^\Omega &\approx& - {\delta_{ij} \over 3 \beta^2} \,
\biggl[11 \, l_0^2 \, \beta \, ({\bf \Omega} {\bf \cdot} {\bf
\Lambda}^{(B)}) \, (1 - \epsilon)
\nonumber \\
&& + \tau_0^2 ({\bf \Omega} {\bf \cdot} \bec{\nabla}) \, (\langle
{\bf u}^2 \rangle + \langle {\bf b}^2 \rangle) \biggr] + {\pi \,
\tau_0^2 \over 7 \sqrt{2} \beta} \, (\Omega_i \nabla_{j}
\nonumber \\
&& + \Omega_j \nabla_{i}) \, (\langle {\bf u}^2 \rangle + \langle
{\bf b}^2 \rangle),
\label{SS15}\\
\bec{\delta}^\Omega &\approx& {17 \pi \, l_0^2 \over 14 \sqrt{2}
\beta} (1 - \epsilon) \, {\bf \Omega} \;,
\label{S19} \\
\kappa_{ijk}^\Omega &\approx& 8 \, {l_0^2 \over \beta} \, (1 -
\epsilon) (\Omega_j \delta_{ik} + \Omega_k \delta_{ij}) \;,
\label{S20}
\end{eqnarray}
$\eta_{ij}^\Omega = O(\Omega^2)$, and we took into account that
$\langle {\bf u}^2 \rangle + \langle {\bf b}^2 \rangle \approx
\langle {\bf u}^2 \rangle^{(0)} + \langle {\bf b}^2 \rangle^{(0)}
 + O(\nabla \bar B)$. Asymptotic formulas (\ref{S12})-(\ref{S14}) and
(\ref{SS12})-(\ref{S18}) in the limit of a very small mean
magnetic field coincide with those obtained in \cite{RKR03} for
$q=5/3$.

Now we study the effect of the mean differential rotation on the
nonlinear electromotive force.  We take into account the tensors
$I_{ijmn}^\sigma(\bar{\bf U})$, $\, E_{ijmn}^\sigma(\bar{\bf U})$
and $J_{ijmn}^\sigma(\bar{\bf U})$ in Eqs.~(\ref{B6})-(\ref{B8}).
The contribution, ${\cal E}^\sigma_{i},$ to the nonlinear
electromotive force caused by a mean velocity shear is determined
by
\begin{eqnarray}
{\cal E}^\sigma_{i} &=& \varepsilon_{inm} \int {\tau \over 1 + 2
\psi} \biggl[J_{mnpq}^\sigma \tilde g_{pq} + i \tau ({\bf k} {\bf
\cdot} \bar{\bf B}) [I_{mnpq}^\sigma \tilde f_{pq}
\nonumber \\
& & + I^{(f,\sigma)}_{mn} - I^{(h,\sigma)}_{mn}] +
I^{(g,\sigma)}_{mn}\biggr] \, d{\bf k} \;  \label{K6}
\end{eqnarray}
(for details, see \cite{RK04}), where the source terms
$I^{(f,\sigma)}_{ij} \equiv I^{f}_{ij}(g_{ij}^{\sigma})$, $\,
I^{(h,\sigma)}_{ij} \equiv I^{h}_{ij}(g_{ij}^{\sigma})$ and
$I^{(g,\sigma)}_{ij} \equiv I^{g}_{ij}(f_{ij}^{\sigma},
h_{ij}^{\sigma})$ are determined by Eqs.~(\ref{MM1})-(\ref{MM3}),
in which $f_{ij}$, $h_{ij}$, $g_{ij}$ are replaced by the
corresponding correlation functions $f_{ij}^{\sigma}$,
$h_{ij}^{\sigma}$, $g_{ij}^{\sigma}$ that describe the
contributions caused by a mean velocity shear. After the
integration in Eq.~(\ref{K6}), we obtain
\begin{eqnarray}
{\cal E}^\sigma_{i} = a_{ij}^{\sigma} \bar B_{j} +
b_{ijk}^{\sigma} \bar B_{j,k} \; .
\label{SS8}
\end{eqnarray}
The tensor $a_{ij}^{\sigma}$ for an inhomogeneous turbulence is
given by Eq.~(\ref{K5}) below. For a homogeneous turbulence
$a_{ij}^{\sigma} = 0$. This case has been considered in
\cite{RK04}. The tensor $b_{ijk}^{\sigma}$ is given by
\begin{eqnarray}
b_{ijk}^{\sigma} = l_0^2 \, \biggl[\sum_{n=1}^7 Q_n \,
S_{ijk}^{(n)} \biggr] \; \label{SS9}
\end{eqnarray}
(see \cite{RK04}), where the coefficient $ Q_3 = 0 $, and the
other coefficients calculated for $q=5/3$ are given by
\begin{widetext}
\begin{eqnarray*}
Q_1 &=& {1 \over 3} \biggl[A_1^{(2)} - 3 A_2^{(2)} - 18 C_1^{(2)}
+ \epsilon \biggl(A_1^{(2)} + A_2^{(2)} + {2 \over 3} C_1^{(2)}
\biggr) + \tilde \Psi_1 \biggl\{A_1 + 2 A_2 + {34 \over 3} C_1 -
\epsilon \biggl(2 A_1 + A_2 + {10 \over 3} C_1\biggr) \biggr\}
\\
&& + \tilde \Psi_2 \biggl\{- A_1 + {7 \over 3} C_1 + \epsilon (A_1
- 5 C_1) \biggr\} - (1 - \epsilon) \tilde \Psi_3\{C_1\} - \tilde
\Psi_0\{2 A_1 - 3 C_1\} \biggr] \;,
\\
Q_2 &=& {1 \over 3} \biggl[-(A_1^{(2)} + A_2^{(2)} + 4 C_1^{(2)})
+ \epsilon \biggl(- A_1^{(2)} + A_2^{(2)} + {32 \over 3} C_1^{(2)}
\biggr) + \tilde \Psi_1 \biggl\{- A_1 + A_2 + {74 \over 3} C_1 - 2
\epsilon \biggl(A_2 + {61 \over 3} C_1\biggr) \biggr\}
\\
&& + \tilde \Psi_2 \{A_1 - 27 C_1 - \epsilon (A_1 - 35 C_1)\} + (1
- \epsilon) \biggl(\tilde \Psi_3\{- 2 A_1 + 7 C_1\} - {64 \over 3}
\tilde \Psi_4\{C_1\} + 16 \tilde \Psi_5\{C_1\} \biggr)
\\
&& + \tilde \Psi_0 \biggl\{2 A_1 - {11 \over 3} C_1 \biggr\}
\biggr] \;,
\\
Q_4 &=& {1 \over 6} \biggl[3 A_1^{(2)} + A_2^{(2)} - {14 \over 3}
C_1^{(2)} + \epsilon \biggl(3 A_1^{(2)} - A_2^{(2)} - {26 \over 3}
C_1^{(2)} \biggr) - \tilde \Psi_1 \biggl\{A_1 + A_2 - {8 \over 3}
C_1 - 2 \epsilon \biggl(A_1 + A_2
\\
&& + {4 \over 3} C_1 \biggr) \biggr\} + (1 - \epsilon)
\biggl(\tilde \Psi_2\{A_1 + C_1\} - \tilde \Psi_3\{C_1\} \biggr) +
\tilde \Psi_0 \{C_1\} \biggr] \;,
\\
Q_5 &=& {1 \over 6} \biggl[A_1^{(2)} + A_2^{(2)} - {14 \over 3}
C_1^{(2)} + \epsilon \biggl(A_1^{(2)} - A_2^{(2)} - {26 \over 3}
C_1^{(2)} \biggr) - \tilde \Psi_1 \biggl\{A_1 - A_2 - {8 \over 3}
C_1 - 2 \epsilon \biggl(A_1 - A_2
\\
&& + {4 \over 3} C_1 \biggr) \biggr\} + (1 - \epsilon)
\biggl(\tilde \Psi_2\{A_1 + C_1\} - \tilde \Psi_3\{C_1\} \biggr) +
\tilde \Psi_0 \{C_1\} \biggr] \;,
\\
Q_6 &=& {1 \over 3} \biggl[A_2^{(2)} - 4 C_3^{(2)} - \epsilon
\biggl(A_2^{(2)} - {32 \over 3} C_3^{(2)} \biggr) + \tilde \Psi_1
\biggl\{- 3 A_2 + {74 \over 3} C_3 + 2 \epsilon \biggl(A_2 - {61
\over 3} C_3\biggr) \biggr\} - (27 - 35 \epsilon) \tilde \Psi_2
\{C_3\}
\\
&& - (1 - \epsilon) \biggl(\tilde \Psi_3\{A_2 - 7 C_3\} + {64
\over 3} \tilde \Psi_4\{C_3\} - 16 \tilde \Psi_5\{C_3\} \biggr) +
\tilde \Psi_0 \biggl\{A_2 - {11 \over 3} C_3 \biggr\} \biggr] \;,
\\
Q_7 &=& {1 \over 6} \biggl[A_2^{(2)} - {14 \over 3} C_3^{(2)} +
\epsilon \biggl(3 A_2^{(2)} - {26 \over 3} C_3^{(2)} \biggr) +
\tilde \Psi_1 \biggl\{A_2 + {8 \over 3} (1 + \epsilon) C_3\biggr\}
+ (1 - \epsilon) \biggl(\tilde \Psi_2 \{2 A_2 + C_3\}
\\
&& - \tilde \Psi_3\{A_2 + C_3\}\biggr) + \tilde \Psi_0 \{A_2 +
C_3\} \biggr] \; .
\end{eqnarray*}
\end{widetext}
\noindent
Here
\begin{eqnarray*}
S_{ijk}^{(1)} &=& \varepsilon_{ijp} (\partial \bar U)_{pk} \;,
\quad S_{ijk}^{(2)} = \varepsilon_{ikp} (\partial \bar U)_{pj} \;,
\\
S_{ijk}^{(3)} &=& \varepsilon_{jkp} (\partial \bar U)_{pi} \;,
\quad S_{ijk}^{(4)} = \bar W_{k} \delta_{ij} \;, \quad
S_{ijk}^{(5)} = \bar W_{j} \delta_{ik} \;,
\\
S_{ijk}^{(6)} &=& \varepsilon_{ikp} \beta_{jq} (\partial \bar
U)_{pq} \;, \quad S_{ijk}^{(7)} = \bar W_{k} \beta_{ij} \; .
\end{eqnarray*}
The coefficients defining the shear-current effect and the
nonhelical $\alpha$ effect are determined by
\begin{eqnarray}
\sigma_0 &=& {1 \over 2} (Q_2 + 2 Q_4 + Q_6 + 2 Q_7) \;,
\label{S30}\\
\sigma_1 &=& - \sigma_0 - {1 \over 2} (Q_1 + 2 Q_5) \;,
\label{S31}
\end{eqnarray}
Thus, the nonlinear coefficient $\sigma_0(\bar{B})$ and
$\sigma_1(\bar{B})$ are determined by
\begin{eqnarray}
\sigma_0(\bar{B}) &=& \Psi_a\{A_1 + A_2\} + \Psi_b\{C_1 + C_3\}
\;,
\label{F1}\\
\sigma_1(\bar{B}) &=& - \sigma_0(\bar{B}) + \Psi_c\{A_1\} +
\Psi_d\{A_2\} + \Psi_e\{C_1\}  \;,
\nonumber\\
\label{F5}
\end{eqnarray}
where
\begin{eqnarray*}
\Psi_a\{X\} &=& {1 \over 3} \biggl[(1 + \epsilon)
X^{(2)}(\sqrt{2}\beta) + [\tilde \Psi_0 - (1 - \epsilon)(\tilde
\Psi_1
\\
&& - \tilde \Psi_2 + \tilde \Psi_3)]\{X\} \biggr] \;,
\\
\Psi_b\{X\} &=& {1 \over 9} \biggl[(3 \epsilon - 13)
X^{(2)}(\sqrt{2}\beta) + [12 \tilde \Psi_2 - 4\tilde \Psi_0
\\
&& - 16 \tilde \Psi_1 + (1 - \epsilon)(57 \tilde \Psi_1 - 51
\tilde \Psi_2 + 9 \tilde \Psi_3
\\
&& - 32 \tilde \Psi_4 + 24 \tilde \Psi_5)]\{X\} \biggr] \;,
\\
\Psi_c\{X\} &=& {1 \over 3} \biggl[- (1 + \epsilon)
X^{(2)}(\sqrt{2}\beta) + \tilde \Psi_0\{X\} \biggr] \;,
\\
\Psi_d\{X\} &=& {1 \over 6} \biggl[2 X^{(2)}(\sqrt{2}\beta) - 3 (1
- \epsilon) \tilde \Psi_1\{X\} \biggr] \;,
\\
\Psi_e\{X\} &=& {1 \over 9} \biggl[(34 + 12 \epsilon)
X^{(2)}(\sqrt{2}\beta) + [4 \tilde \Psi_2 - 6 \tilde \Psi_0
\\
&& - 20 \tilde \Psi_1 - (1 - \epsilon)(\tilde \Psi_1 + 9 \tilde
\Psi_2 - 3 \tilde \Psi_3)]\{X\} \biggr] \;,
\end{eqnarray*}
and the functions $\tilde \Psi_k\{X\}$ are given by
\begin{eqnarray}
\tilde \Psi_0\{X\} &=& - {1 \over 2} (1 + \epsilon) X^{(2)}(0) +
(2 - \epsilon) X^{(2)}(\sqrt{2}\beta)
\nonumber\\
&& - {3 \over 4 \pi} (1 - \epsilon) \bar X(2 \beta^2) \;,
\nonumber \\
\tilde \Psi_1\{X\} &=& - 3 X^{(2)}(\sqrt{2}\beta) + {3 \over 2
\pi} \bar X(2 \beta^2) \;,
\nonumber \\
\tilde \Psi_2\{X\} &=& 3 X^{(2)}(\sqrt{2}\beta) - {3 \over 2 \pi}
\biggl[\bar X(y) + {1 \over 2} \, y \bar X'(y)
\biggr]_{y=2\beta^2} \;,
\nonumber\\
\tilde \Psi_3\{X\} &=& - 6 X^{(2)}(\sqrt{2}\beta) + {3 \over 2
\pi} \biggl[2 \bar X(y)
\nonumber\\
&& + {1 \over 2} \, y \bar X'(y) \biggr]_{y=2\beta^2} \;,
\nonumber\\
\tilde \Psi_4\{X\} &=& 4 X^{(2)}(\sqrt{2}\beta) - {1 \over \pi}
\biggl[2 \bar X(y) + y \bar X'(y)
\nonumber\\
&& + {1 \over 4} y^2 \bar X''(y) \biggr]_{y=2\beta^2} \;,
\nonumber\\
\tilde \Psi_5\{X\} &=& - {1 \over 2} X^{(2)}(\sqrt{2}\beta) + {1
\over 4 \pi} \biggl[\bar X(y)
\nonumber\\
&& + {1 \over 2} \, y \bar X'(y) + y^2 \bar X''(y)
\biggr]_{y=2\beta^2} \; . \label{G1}
\end{eqnarray}

The tensor $a_{ij}^{\sigma}$ is given by
\begin{eqnarray}
a_{ij}^{\sigma} &=& - {l_0^2 \over 6} \biggl[F_{1} \, \delta_{ij}
\, (\bar{\bf W} {\bf \cdot} {\bf \Lambda}^{(v)}) + F_{2} \,
\bar{W}_i \Lambda^{(v)}_j + F_{3} \, \bar{W}_j \Lambda^{(v)}_i
\nonumber\\
&& + F_{4} \, S_{ijn}^{(1)} \Lambda^{(v)}_n + F_{5} \,
S_{ijn}^{(2)} \Lambda^{(v)}_n + \epsilon [F_{6} \, \delta_{ij} \,
(\bar{\bf W} {\bf \cdot} {\bf \Lambda}^{(b)})
\nonumber\\
&& + F_{7} \, \bar{W}_i \Lambda^{(b)}_j + F_{8} \, \bar{W}_j
\Lambda^{(b)}_i + F_{9} \, S_{ijn}^{(1)} \Lambda^{(b)}_n
\nonumber\\
&& + F_{10} \, S_{ijn}^{(2)} \Lambda^{(b)}_n] \biggr]\;,
\label{K5}
\end{eqnarray}
where
\begin{eqnarray*}
F_{1} &=& (3 G^{(2)} - H^{(2)})\{A_1\} + {1 \over 2}
[H^{(2)}\{A_2\}
\\
&& + A_2^{(2)}(\sqrt{2}\beta)] \;,
\\
F_{2} &=& - {1 \over 2} [(6 G^{(2)} - 3 H^{(2)})\{A_1\} -
A_1^{(2)}(\sqrt{2}\beta)] \;,
\\
F_{3} &=& - {1 \over 2} [H^{(2)}\{A_1 + A_2\} +
A_1^{(2)}(\sqrt{2}\beta)
\\
&& + A_2^{(2)}(\sqrt{2}\beta)] \;, \quad F_{4} = - 2 F_{3} \;,
\quad F_{9} = - 2 F_{8} \;,
\\
F_{5} &=& - 2 F_{1} + {4 \over 3} \biggl[(3 G^{(2)} + 8
H^{(2)})\{C_1 + C_3\}
\\
&& + 4 [C_1^{(2)}(\sqrt{2}\beta) + C_3^{(2)}(\sqrt{2}\beta)]
\biggr] \;,
\\
F_{6} &=& - 3 (G^{(2)} - H^{(2)})\{A_1\} - {1 \over 2}
[H^{(2)}\{A_2\}
\\
&& - A_2^{(2)}(\sqrt{2}\beta)] \;,
\\
F_{7} &=& {1 \over 2} [(6 G^{(2)} - 7 H^{(2)})\{A_1\} +
A_1^{(2)}(\sqrt{2}\beta)] \;,
\\
F_{8} &=& {1 \over 2} [H^{(2)}\{A_1 + A_2\} -
A_1^{(2)}(\sqrt{2}\beta) - A_2^{(2)}(\sqrt{2}\beta)] \;,
\\
F_{10} &=& - 2 F_{6} -  4 (G^{(2)} + 2 H^{(2)})\{C_1 + C_3\}
\\
&& + {16 \over 3} [C_1^{(2)}(\sqrt{2}\beta) +
C_3^{(2)}(\sqrt{2}\beta)] \;,
\end{eqnarray*}
and
\begin{eqnarray*}
G^{(2)}\{X\} &=&  10 X^{(2)}(\sqrt{2}\beta) - {3 \over 4 \pi} [6
\bar X(y)
\\
&& + y \bar X'(y)]_{y=2\beta^2} \;,
\\
H^{(2)}\{X\} &=&  4 X^{(2)}(\sqrt{2}\beta) - {3 \over 2 \pi} \bar
X(2\beta^2)
\\
&=& \Psi_2\{X\}_{\sqrt{2}\beta}\;,
\\
G^{(2)}\{X\} &-& H^{(2)}\{X\} = \Psi_3\{X\}_{\sqrt{2}\beta} \;,
\end{eqnarray*}
For the derivation of Eq.~(\ref{K5}) we used the following
identities
\begin{eqnarray*}
&& \varepsilon_{inm} \Lambda_n \bar B_{j} \bar K_{jmpq} \nabla_p
\bar U_q = 2(\bar C_1 + \bar C_3) \varepsilon_{inq} \Lambda_n
(\partial \bar U)_{qj} \bar B_{j} \;,
\\
&& (\varepsilon_{imq} \bar K_{jm} \delta_{pn} - \varepsilon_{inq}
\bar K_{jp}) \Lambda_n \bar B_{j} \nabla_p \bar U_q = (\bar A_1 +
\bar A_2) \Lambda_p \bar B_{j}
\\
&& \times \biggl[\varepsilon_{ijq} (\partial \bar U)_{pq} -
\varepsilon_{ipq} (\partial \bar U)_{jq} - {1 \over 2}
(\delta_{ip} \bar{W}_j + \delta_{ij} \bar{W}_p) \biggr] \;,
\\
&& \varepsilon_{inm} \bar K_{mq} \Lambda_n \bar B_{j} \nabla_p
\bar U_q = \bar A_1 \biggl[\varepsilon_{inq} (\partial \bar
U)_{jq} + {1 \over 2} (\delta_{jn} \bar{W}_i
\\
&& - \delta_{ij} \bar{W}_n) \biggr] \Lambda_n \bar B_{j} \; .
\end{eqnarray*}
An additional contribution to the isotropic part $(\alpha_{ij}
\propto \alpha \delta_{ij})$ of the nonlinear $\alpha$ effect [see
Eq.~(\ref{C4})] due to both, inhomogeneity of turbulence and mean
differential rotation in a nondimensional form in spherical
coordinates is given by
\begin{eqnarray}
\alpha^{\delta \Omega} &=& {L \, W_\ast \over L_{_{T}}} \,
\biggr[\tilde \Psi_6\{X\} \Lambda^{(v)} + \epsilon \tilde
\Psi_7\{X\} \Lambda^{(b)}\biggr]_{X=C_1 + C_3}
\nonumber \\
&& \times \sin \theta {\partial \over
\partial \theta}(\delta \Omega) \;, \label{K8}
\end{eqnarray}
where
\begin{eqnarray*}
\tilde \Psi_6\{X\} &=& - {22 \over 3} X^{(2)}(\sqrt{2}\beta) + {1
\over 12 \pi} \biggl[34 \bar X(y)
\nonumber \\
&& + y \bar X'(y)\biggr]_{y=2\beta^2} \;,
\\
\tilde \Psi_7\{X\} &=& {50 \over 9} X^{(2)}(\sqrt{2}\beta) - {1
\over 4 \pi} \biggl[10 \bar X(y)
\nonumber \\
&& + y \bar X'(y)\biggr]_{y=2\beta^2} \; .
\end{eqnarray*}
The contribution to the nonlinear $\alpha$ effect due to both,
inhomogeneity of turbulence and mean differential rotation for a
weak mean magnetic field $\bar{B} \ll \bar{B}_{\rm eq} / 4$ is
given by
\begin{eqnarray}
\alpha^{\delta \Omega} &=& - {2 \over 9} \,  {L \, W_\ast \over
L_{_{T}}} \, \biggl[\Lambda^{(v)} - {\epsilon \over 3}
\Lambda^{(b)}\biggr] \sin \theta {\partial \over \partial
\theta}(\delta \Omega) \;,
\nonumber \\
\label{K9}
\end{eqnarray}
and for $\bar{B} \gg \bar{B}_{\rm eq} / 4$ it is given by
\begin{eqnarray}
\alpha^{\delta \Omega} &=& - {1 \over 9 \beta^2} \, {L \, W_\ast
\over L_{_{T}}} \, \epsilon \Lambda^{(b)} \sin \theta {\partial
\over \partial \theta}(\delta \Omega) \; . \label{K10}
\end{eqnarray}
Equations for $\alpha^{\delta \Omega}$ in cylindrical coordinates
can be obtained from Eqs.~(\ref{K8})-(\ref{K10}) after the change
$\sin \theta (\partial / \partial \theta) \to \rho (\partial /
\partial \rho)$.

Note that the $\alpha^{\delta \Omega}$ term has been also calculated
in \cite{RS05} for a kinematic problem using the second-order
correlation approximation (SOCA).

\end{document}